\documentclass[galaxies,review,accept,pdftex,moreauthors]{Definitions/mdpi} 

\DeclareUnicodeCharacter{2212}{-}

\firstpage{1} 
\pubvolume{13}
\issuenum{5}
\articlenumber{99}
\pubyear{2025}
\copyrightyear{2025}
\externaleditor{Jorick S. Vink} 
\datereceived{10 June 2025} 
\daterevised{19 August 2025} 
\dateaccepted{20 August 2025} 
\datepublished{1 September 2025} 
\hreflink{https://doi.org/10.3390/\linebreak galaxies13050099} 

\usepackage{url}
\usepackage{multirow}
\usepackage{longtable} 
\Title{The Astronomical Hub: A Unified Ecosystem for Modern Astronomical Research $^{\dagger}$}

\TitleCitation{The Astronomical Hub: A Unified Ecosystem for Modern Astronomical Research}

\Author{Yerlan 
 Aimuratov 
 *\orcidA{}, 
        Vitaliy Kim *\orcidB{},
        Aleksander Serebryanskiy~\orcidC{},
        Denis Yurin~\orcidD{},  
        Maxim Krugov~\orcidE{},
        Chingiz~Akniyazov~\orcidF{}, 
        Saule Shomshekova~\orcidG{}, 
        Maxim Makukov~\orcidH{}, 
        Gaukhar Aimanova~\orcidI{}, 
        Rashit Valiullin~\orcidJ{},
        Raushan Kokumbaeva~\orcidK{},
        Alan Kazkenov  and Chingis Omarov~\orcidL{} 
        }

\AuthorNames{Yerlan Aimuratov, Vitaliy Kim, Aleksander Serebryanskiy, Denis Yurin, Maxim Krugov, Chingiz Akniyazov, Saule Shomshekova, Maxim Makukov, Gaukhar Aimanova, Rashit Valiullin, Raushan Kokumbaeva, Alan Kazkenov and Chingis Omarov}

\isAPAStyle{%
        \AuthorCitation{Aimuratov, Y., Kim, V., Serebryanskiy, A., Yurin, D., Krugov M., Akniyazov C., Shomshekova, S., Makukov, M., Aimanova, G., Valiullin, R., Kokumbaeva R., Kazkenov, A., \& Omarov, C.}
         }{%
        \isChicagoStyle{%
        \AuthorCitation{Yerlan Aimuratov, Vitaliy Kim, Aleksander Serebryanskiy, Denis Yurin, Maxim Krugov, Chingiz Akniyazov, Saule Shomshekova, Maxim Makukov, Gaukhar Aimanova, Rashit Valiullin, Raushan Kokumbaeva, Alan Kazkenov and Chingis Omarov}
        }{
        \AuthorCitation{Aimuratov, Y.; Kim, V.; Serebryanskiy, A.; Yurin, D.; Krugov, M.; Akniyazov, C.; Shomshekova, S.; Makukov, M.; Aimanova, G.; Valiullin, R.; et al.}
        }
}

\address[1]{%
Fesenkov Astrophysical Institute, Observatory 23, Medeu District, Almaty 050020, Kazakhstan
\\
}

\corres{\hangafter=1 \hangindent=1.05em \hspace{-0.82em}Correspondence: aimuratov@fai.kz (Y.A.); kim@fai.kz (V.K.)}

\firstnote{\hangafter=1 \hangindent=1.05em \hspace{-0.82em}Based on a plenary opening talk given by D.Y. at the ``Hot Stars---Life with Circumstellar Matter'' conference, held at the Al-Farabi Kazakh National University on 14--19 October 2024, Almaty, Kazakhstan.}

\abstract{We present the conceptual framework of the Astronomical Hub (AstroHub), a unified platform combining various optical instruments at a single observatory. Its major approach lies in arranging conditions for research groups to install telescopes and equipment and participate in joint projects. AstroHub is planned to integrate Virtual Observatory (VO) tools, FAIR data principles, and a telescope network to create a powerful and attractive ecosystem for both robust near-Earth object (NEO) monitoring and diverse deep space research. We provide an overview of the AstroHub development directions in the case study of the Assy-Turgen Observatory.}

\keyword{telescopes; wide-field telescopes; ground-based astronomy; astronomical site~protection}

\begin{document}
\section{Introduction}\label{sec:introduction}
An integral part of the search for breakthrough areas in science is the productive exchange of ideas, data, and~knowledge. To~this end, technology parks and scientific and educational centres of competence and excellence are being created.\endnote{International Astronomical Center: \url{http://www.astronomycenter.net/} (accessed on 15 July 2025). Centre of Excellence for All-Sky Astrophysics (CAASTRO): \url{https://rsaa.anu.edu.au/about/partnerships/caastro-centre-excellence-all-sky-astrophysics/} (accessed on 15 July 2025).} Astronomical science in this sense is not an exception~\cite{Sanchez...2021rams.book.....S}.

One of the most pivotal aspects of ground-based observatories is international cooperation, which optimises the use of technical resources, provides and expands observational opportunities, promotes scientific exchange, and~addresses complex scientific questions more effectively \citep{NAP12951, 2021pdaa.book.....N, 10.1093/oso/9780198881193.003.0009,2020arXiv200705546C}. The~most advanced ground-based observatories are the centres for the pooling of scientists and engineers that leads to innovative discoveries and breakthroughs in astronomy. Today this also becomes relevant from a scientific perspective, because~over the next decade observational astronomy will face many challenges, ranging from light pollution caused by urbanisation \citep{2017SciA....3E1528K, 2023Sci...379..265K, 2024ARep...68S.236S}, climate change \citep{Cantalloube...2020NatAs...4..826C,Knodlseder...2024arXiv240904054K}, and~exponentially growing space debris \citep{2025MNRAS.541L..47K}, Radio Frequency Interference (RFI), and satellite Internet constellations \citep{2020ApJ...892L..36M, 2020BAAS...52.0206W, 2020A&A...636A.121H,Peel...2025arXiv250411561P} to the~detection and study of dark matter and dark energy. Therefore, for~astronomy, as~for a global science, it becomes crucial to combine intellectual and technical resources in the most advantageous locations for ground-based astronomy in~terms of astroclimate, location, infrastructure, etc. 

The objective of this article is to present an Astronomical Hub (AstroHub), a~set of the best practices unified in an astronomical ecosystem. 
We promote the ongoing development of expertise for ground-based observatories with a case study of the Assy-Turgen Observatory (ATO) that will integrate advanced Kazakhstani and international optical telescopes, near-Earth space monitoring systems, and hardware and software tools for data collection, storage, and~analysis.

The outline of this article is as follows: In Section~\ref{sec:astrohubconcept}, the~concept of the astronomical hub is introduced. We summarise information regarding the levels of integration into the Hub's ecosystem (Section~\ref{sec:integrationlevels}) and integration strategy (Section~\ref{sec:integrationstrategy}). We refer to the importance of interoperability and data management (Section~\ref{sec:interopanddatashare}) for better use of available data and resources. In~Section~\ref{sec:developmentstatus}, we focus on a case study of the Assy-Turgen astronomical hub. We summarise information regarding the integration of diverse telescopic assets, its primary scientific capabilities, particularly in deep space and near-Earth space monitoring, the~underlying technological infrastructure, efforts in capacity building, and~planned future enhancements. In~Section~\ref{sec:researchandcampaigns}, we briefly discuss joint research and international campaigns with an aim to present a coherent overview of this emerging facility for the broader astronomical community. Section~\ref{sec:futureprospects} describes the future prospects of the AstroHub. We give conclusive remarks in Section~\ref{sec:astrohubconclusion}.

\section{The Astronomical~Hub}\label{sec:astrohubconcept}
In this section, we introduce the concept of the Astronomical Hub. We define the philosophy of a unified observatory ecosystem and share our vision of tiered integration levels.  We draw the strategy of integration and mark the milestones that lead to this~goal.

\subsection{The Concept---
A~Unified~Ecosystem}
The Astronomical Hub does not aim to introduce a completely novel astronomical concept. Instead, its purpose is to create a unique and robust synthesis of established and developing best practices. This synthesis will be specifically adapted to the Hub's geographical positioning, research capabilities, and~its local and global objectives. Therefore, AstroHub is more about~technology.

The three core components of this concept are as follows:
\begin{enumerate}[label=,leftmargin=*,labelsep=0mm]
    \item[] \textbf{Platform
}---
The observatory and its associated infrastructure;
    \item[] \textbf{Operator}---The service provider;
    \item[] \textbf{Client}---Individual researchers, research teams, or~research institutions.
\end{enumerate}

\begin{figure}[H]
    \includegraphics[width=0.4\textwidth]{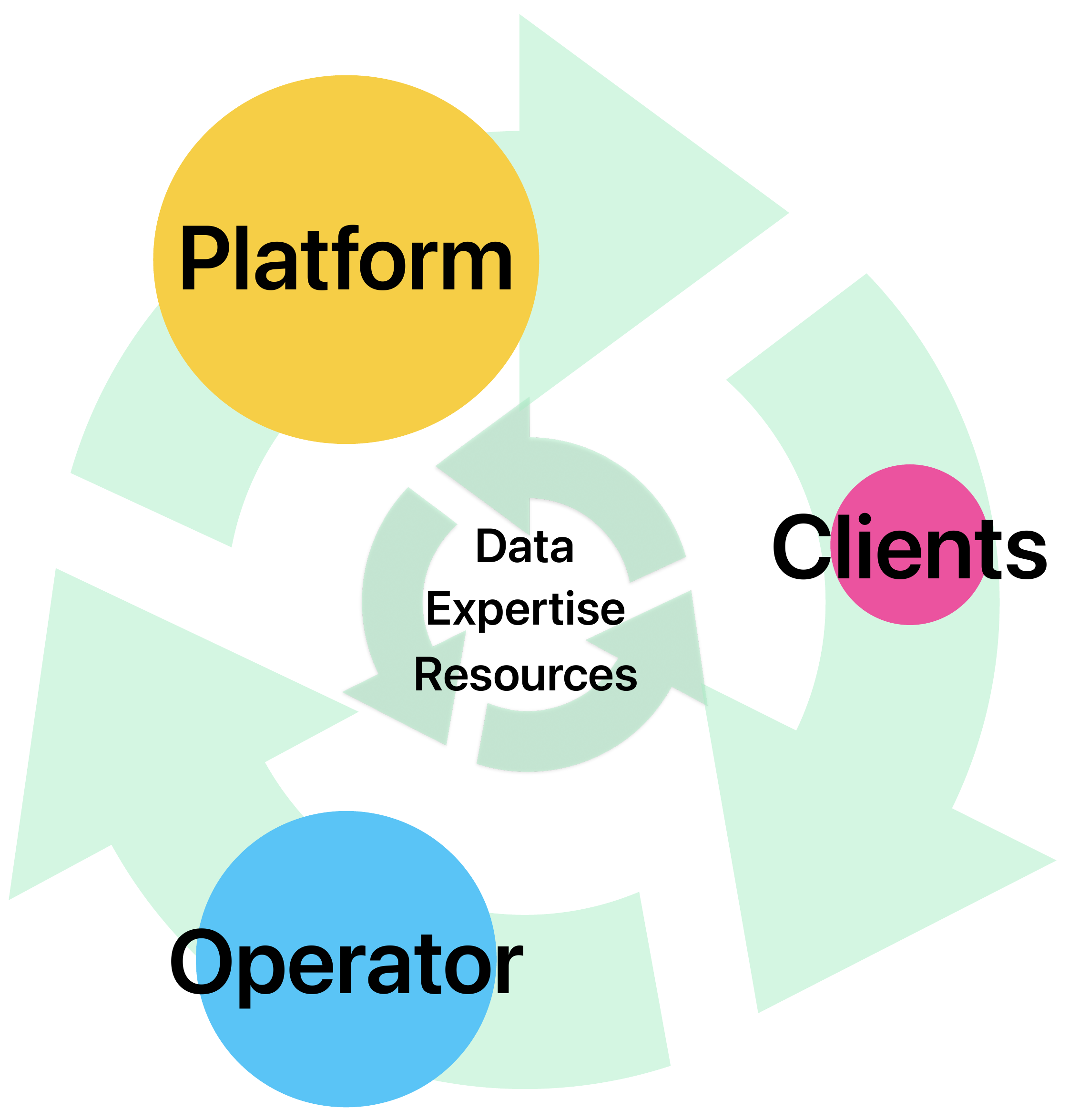}
    \caption{Platform, 
 operator, and~client forming a unit of the AstroHub by continuous exchange of data, resources, and~expertise.}
    \label{fig:astrohubunit}
\end{figure}

These components are interconnected, and~the fundamental principle is the efficient and uninterrupted exchange of data, resources, and~expertise within each unit, see Figure~\ref{fig:astrohubunit}.

\unskip

\subsection{Integration~Levels}\label{sec:integrationlevels}
The practice of deploying an individual or institutional telescope at a host observatory, commonly known as ``telescope hosting'', is a recognised model in astronomical research. This arrangement accommodates a spectrum of users, from~amateur astrophotographers seeking optimised observation conditions to professional research entities deploying specialised instrumentation. Derived from observatory best practices, a~management protocol was established, encompassing key procedural stages and considerations, as~detailed in Appendix \ref{sec:telescopehostingscheme}. These procedures typically involve formalised, legally binding agreements, progressing from a scientific and technical proposal to a comprehensive hosting contract, followed by meticulously supervised installation and a delineated operational phase. The~degree of integration and utilisation of shared services constitutes the primary variable. And~it is precisely this ``integration level'' that has been central to the development of a versatile AstroHub~ecosystem.

\begin{figure}[H]
 \includegraphics[width=0.9\textwidth]{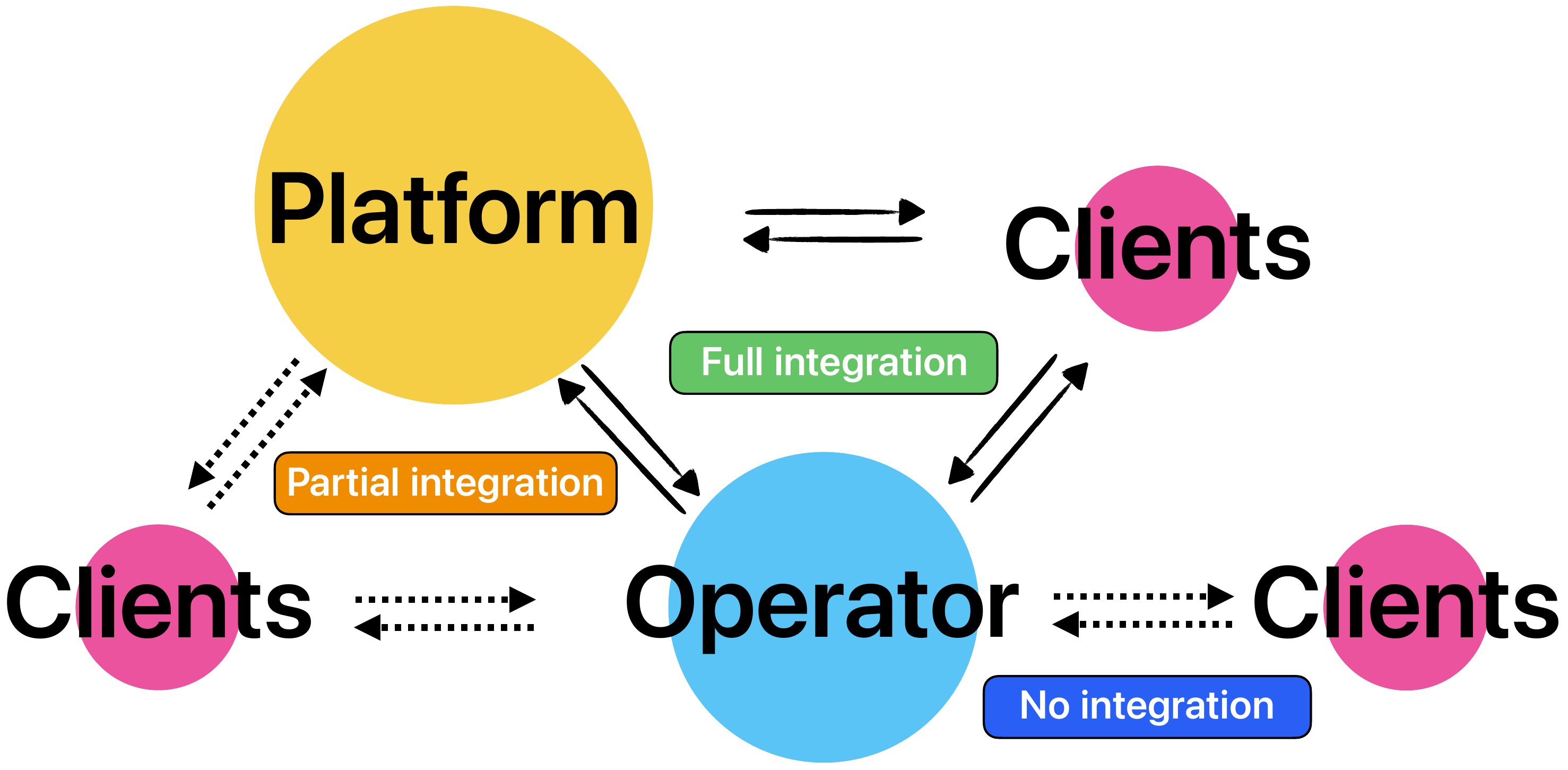}
    \caption{AstroHub integration levels. Three levels---full, partial, and~no integration---define the interaction type within three kinds of units, each consisting of a platform, operator, and~client.}
    \label{fig:astrohubintegrationlevels}
\end{figure}

We have defined three levels of client participation in the AstroHub ecosystem. These levels empower clients to select the optimal interaction approach with both the operator and the platform, aligning with their unique requirements and available resources. The~subsequent development will focus on the following delineated participation \mbox{levels (see Figure~\ref{fig:astrohubintegrationlevels}):}

\begin{enumerate}
    \item Full integration;
    \item Partial integration;
    \item No integration (full autonomy).
\end{enumerate}

This concept of tiered participation levels adds a crucial layer of practical and strategic depth to the AstroHub ecosystem. It is designed to provide a sophisticated understanding of the diverse scientific goals and motivations of potential~collaborators.\\
\vspace{-8pt}

\noindent
\textit{Level 
 1: Full Integration.}
\begin{enumerate}[label=,leftmargin=*,labelsep=0mm]
    \item[]\textbf{Concept: 
} A full partnership model. The~client's integrated telescope becomes a fully fledged node in the AstroHub network.
    \item[]\textbf{Give/Get: 
} The client shares their telescope time with the network and, in~return, gains access to observing time on all other integrated telescopes. This is the full realisation of the unified ecosystem that is our essential basis. 
    \item[]\textbf{Impact: 
} Full integration maximises the use of AstroHub resources as a client obtains access to all services and infrastructure, including data storage, the processing platform, the library of algorithms and catalogues, and~technical support. The~client can share their data and algorithms with other participants and actively participate in community joint projects. When data and algorithms are fully integrated with the AstroHub system, it ensures maximum efficiency and provides a high level of interoperability.
\end{enumerate}

\noindent
\textit{Level 2: Partial Integration.}
\begin{enumerate}[label=,leftmargin=*,labelsep=0mm]
    \item[]\textbf{Concept 
}: A flexible, hybrid, or~``\`a la carte'' model. This is for clients who want some benefits of the ecosystem without full commitment.
    \item[]\textbf{Give/Get: 
} The client and the operator come to a custom agreement. This could involve sharing specific types of data (e.g., only transient alerts), offering a certain percentage of telescope time, or~using specific AstroHub services (such as data processing or archiving) while maintaining autonomy in other areas. At~the same time, the~client will have the opportunity to equip the installed telescope with its own storage, communication, or~software solutions. The~operator, in~turn, has the right to integrate part of its observing time on this telescope as part of a network-wide AstroHub system, providing other clients with the opportunity to utilise the full resources, including this telescope.
    \item[]\textbf{Impact
}: Partial integration implies a selective use of services. The~client can limit access to its data and algorithms, which results in limited participation in the community. The~client can use its data formats and algorithms while also having the ability to integrate with the AstroHub system as needed, which still keeps a moderate level \mbox{of interoperability.}
\end{enumerate}

\noindent
\textit{Level 3: No Integration (Full Autonomy).}
\begin{enumerate}[label=,leftmargin=*,labelsep=0mm]
    \item[]\textbf{Concept}: A hosting model. The~client is physically present but operationally independent.
    \item[]\textbf{Give/Get}: The client receives basic, essential resources (power supply, dome, connectivity, weather information, etc.) but does not share data or telescope time with the AstroHub community, nor do they use the Hub's integrated services. Therefore, their participation in the ecosystem is minimal.
    \item[]\textbf{Impact}: The client rents a site, a~pavilion, receives energy services, and~receives information on weather conditions entirely on a commercial basis. All other aspects related to observations, data processing, and~storage are provided by the client independently; i.e.,~the client will have the opportunity to equip the installed telescope with its own storage, communication, or~software solutions. At~the same time, the~AstroHub platform can additionally provide services for observations, equipment repair, data storage, and~processing. All these ensure minimal interaction: the client contacts the operator only regarding issues related to the infrastructure. While the client does not integrate with the AstroHub system and uses its own solutions for all aspects of work, there is a lack of proper interoperability.
\end{enumerate}

How does this refine the classical telescopes' hosting approach to our AstroHub original idea? The tiered structure makes the AstroHub's operational model much more robust and attractive. It acknowledges that not every researcher has the same goals. When a university might desire \textit{full integration} 
 to boost research collaboration and science connections, a~private company focused on a proprietary NEO survey might prefer \textit{full autonomy} or a very specific negotiated agreement with a clear plan of \textit{partial integration}.

\subsection{Integration~Strategy}\label{sec:integrationstrategy}
Despite potential superficial similarities and varied involvement levels, including traditional telescope hosting or extensive scientific collaboration, the~AstroHub model possesses distinct structural organisation and unique~benefits.

\begin{figure}[H]
    \hspace{-12pt}\includegraphics[width=\textwidth]{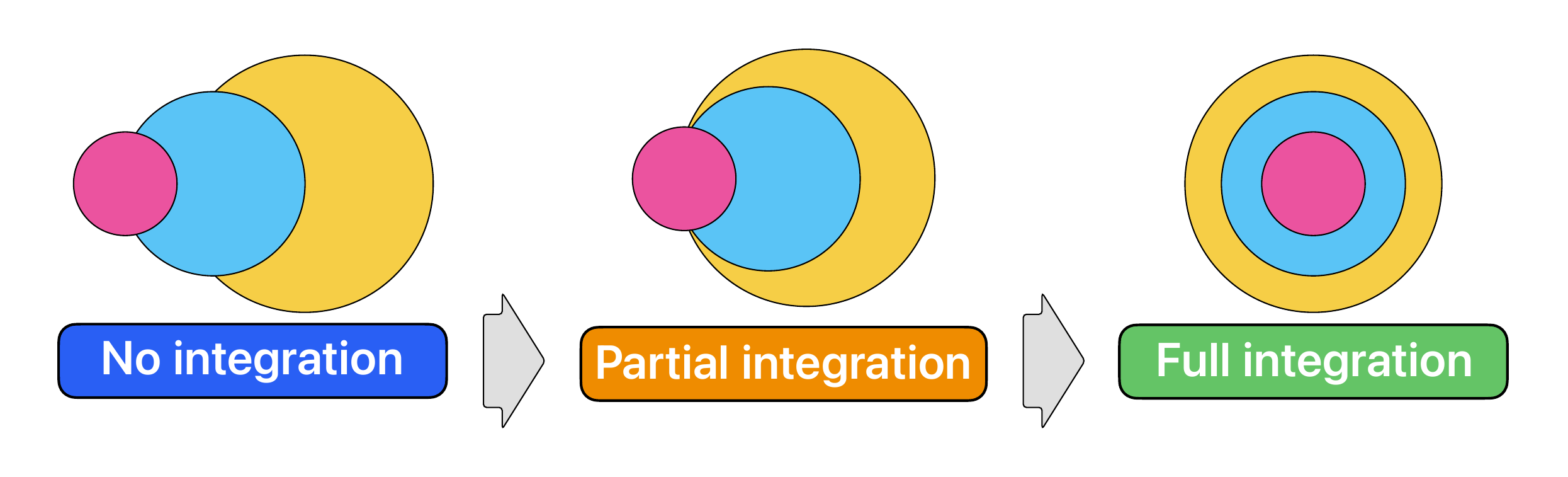}
    \caption{AstroHub's ``Integration by Design'' model. Circles represent platform (yellow), operator (blue), and~client (pink), with three integration levels progressively increasing the client's interaction with operator and deepening it into the~platform.}
    \label{fig:astrohubintegrationupgrade}
\end{figure}

The usual scenario is the most common form of commercial telescope hosting.\endnote{Sierra Remote Observatories: \url{https://www.sierra-remote.com}. Starfront Observatories: \url{https://starfront.space}.} An observatory rents out a piece of land and provides power and~Internet. Here, the~difference is subtle but important. The~advantage of AstroHub, even for individual researchers, lies in its environment and the potential for future collaboration. They are setting up their equipment in a dynamic, growing ecosystem with a clear, structured path to upgrade their participation level if their scientific goals change, see Figure~\ref{fig:astrohubintegrationupgrade}. Therefore, we can underline it as ``Integration by Design'', and~this is where the AstroHub model starts to diverge and show its strengths, especially when we look at its \textit{full integration} and \mbox{\textit{partial integration} levels.}

\vspace{-9pt}

The hosting model of collaboration often happens in a bespoke, ad hoc manner. A~university might negotiate a one-off deal with a national observatory. A~consortium might form to build a single, massive telescope. These are often complex, multi-year negotiations that result in a rigid structure tailored to that one specific project. AstroHub's model and platform were deliberately designed to ensure flexibility, scalability, and~\mbox{synergistic functionality.}

The \textit{partial integration} model formalises a unique agreement to find a solution that sits between full autonomy and a deep partnership. It essentially creates a ``service catalogue'' menu. A~new client can keep her/his own scheduling but~wants to use the data processing pipeline and archiving service. And~in return, e.g.,~the client can offer 15\% of telescope time on targets of opportunity. This is a clean, efficient, and~transparent negotiation based on a pre-defined framework, which turns a complex problem into a manageable discussion. This structured flexibility is a massive advantage that the classic approach simply does not offer~systematically.

AstroHub's \textit{full integration} offers a level of synergy that is rare outside of the world's largest single-project observatories. Even in a collaborative observatory, different hosted telescopes can operate as isolated units. They can have various software, distinct data policies, and~different teams. The~AstroHub's advantage is in promoting a unified ecosystem, and~when a client's telescope is fully integrated, it becomes a node in a larger, \mbox{intelligent network.}

In essence, the~fundamental innovation we propose is lowering barriers to entry. 
 Prospective research groups or individuals who are unsure about committing to full sharing can start with a lower level of integration and potentially upgrade later. And~this is the very moment when the role of the AstroHub operator is of high importance, since they act as service managers, tailoring agreements to fit community~needs.

\subsection{Interoperability of the Telescope~Network}\label{sec:interopanddatashare}
Integrating a new client's telescope or instrument into a distributed network is a profound challenge that goes far beyond physical installation \citep{Sybilski...2014SPIE.9152E..1CS}. This lack of inherent interoperability means that coordinating observations, sharing data in real time, or~managing the network as a single entity becomes a complex, manual, and~inefficient process, reducing the entire network to a group of isolated instruments that happen to share the same~location.

Overcoming this challenge by implementing a robust interoperability system, however, unlocks transformative scientific and operational capabilities \citep{Brown...2013PASP..125.1031B}. It elevates the network from a simple collection of telescopes into a single, cohesive scientific instrument where each element can be made to play its role. 
This synergy enables powerful new modes of observation, such as a wide-field telescope detecting a transient event and automatically triggering a spectroscopic follow-up on another instrument within~minutes.

\begin{figure}[H]
    \begin{adjustwidth}{-\extralength}{0cm}
        \includegraphics[width=\linewidth]{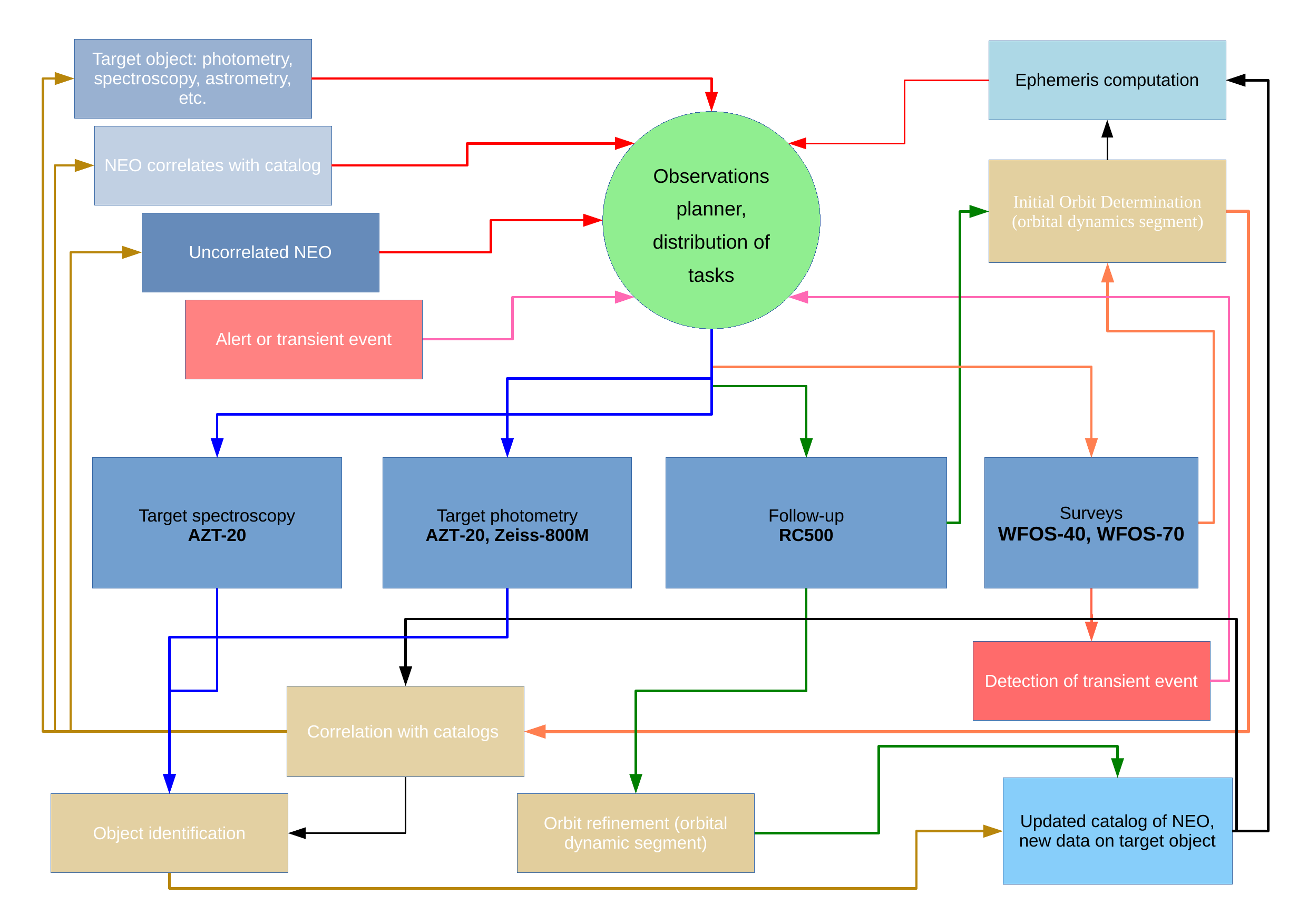}
       \begin{adjustwidth}{+\extralength}{0cm}
 \caption{A scheme of the telescope network interoperability within AstroHub at the Assy-Turgen Observatory.} \label{fig:astrohubinteroperability}
\end{adjustwidth}
       
    \end{adjustwidth}
\end{figure}

The~AstroHub can potentially perform science that no single telescope could. In Figure~\ref{fig:astrohubinteroperability}, we show an example of a pipeline for the alert/scheduled observations of unidentified transient objects. It outlines a scheme for planning observations of various types of objects, including near-Earth objects (NEOs) that are correlated and uncorrelated with catalogues, as~well as transient event alerts. The~process includes (using at the moment exclusively FAI network telescopes for NEOs): 
\begin{itemize}
    \item Object identification and correlation with catalogues;
    \item Detection of transient events;
    \item Ephemeris computation and Initial Orbit Determination;
    \item Observation planning and task distribution;
    \item Observations using various instruments:  
 target spectroscopy, target photometry, follow-up observations, and surveys;
    \item Orbit refinement;
    \item Updating the NEO catalogue and target object data.
\end{itemize}

\vspace{-13pt}

\vspace{-6pt}

In the above pipeline, an~example of an unidentified transient object detected by survey telescope rapid pre-processing can make the central AstroHub system automatically trigger follow-up observations on another telescope, all within minutes or even faster. This is a capability that emerges directly from the ecosystem model. Operational efficiency follows, as everyone benefits from the best practices, software tools, e.g.,~data reduction pipelines and schedulers, and~the expertise of the entire~system. 

\subsection{FAIR Principles and VO~Standards}\label{sec:fairandvo}
Ultimately, interoperability maximises the scientific return on every photon collected across the entire network by enabling unified scheduling, automated data processing pipelines, and~coordinated, multi-telescope observing strategies that would be impossible to achieve otherwise. All these benefits at first also stand as a challenge for every scalable system but, at~the same time, act as a roadmap for coordinated~action.

We envisage implementing and following the FAIR principles \citep{wilkinson2016fair} for the AstroHub. The~FAIR (which stands for ``Findable'', ``Accessible'', ``Interoperable'', and~``Reusable'') principles are a set of guidelines designed to make scientific data more valuable and reusable for both humans and computers \citep{lamprecht2020towards}.

In general, astronomy is a natural fit for the FAIR principles \citep{o2022fair}; in many ways, the~field was practising elements of FAIR long before the acronym was coined. The~tradition of creating data archives and sharing observations \citep{Wenger...2000AAS..143....9W,Taylor...2005ASPC..347...29T,macmillan2012sao} has made astronomy a leading example of how these principles can accelerate~science.

Focused on data, FAIR principles also tune and shape the environment that will obtain that data. Therefore, in~the case of AstroHub, the~FAIR principles can be used for resolving the following issues.

\vspace{3pt}

\noindent
\textit{1. Uniform 
 software and technologies for building the FAIR infrastructure.} 
\begin{enumerate}[label=,leftmargin=*,labelsep=0mm]
    \item[] \textbf{Metadata Management and Database.} 
 The central metadata catalogue is a powerful, open-source (preferably), and~reliable relational database. It should support various data types for storing flexible, instrument-specific metadata alongside the core, standardised~metadata.
    \item[]\textbf{Data Processing and Pipelines.} Today, Python is the standard for astronomical data processing. Astropy \citep{robitaille2013astropy} is a core Python library that provides fundamental tools for astronomy, including robust FITS \citep{grosbol1991fits} file handling, WCS (World Coordinate \mbox{System) \citep{greisen2002representations}} transformations, and~unit conversions. Containerisation ensures that every time a piece of data is processed, it is performed in the exact same software environment, which is crucial for reproducibility.
    \item[]\textbf{Data Access and APIs.} The most standard and well-understood way to provide data is to build web services. Use of International Virtual Observatory Alliance (IVOA) standards~\citep{Berriman...2024ASPC..535..287B} makes data become truly interoperable with the global astronomy community: Table Access Protocol (TAP) \citep{nandrekar2014table}, Simple Image Access (SIA) \citep{dowler2016ivoa}, \mbox{etc.}
\end{enumerate}

\noindent
\textit{2. The~challenges of implementing FAIR in a multi-user environment.}
\begin{enumerate}[label=,leftmargin=*,labelsep=0mm]
    \item[]\textbf{Technical Challenge: Heterogeneity.}  The biggest technical hurdle is the sheer diversity of instruments. Each new telescope from a ``full'' or ``partial'' integration client comes with its own unique hardware, software, and~data quirks. 
   There is a need for a single, mandatory pipeline together with a ``validator'' service that can ensure that the incoming data meets the Hub's minimum standards before it enters the \mbox{main pipeline.}
\end{enumerate}

\noindent
\textit{3. Drafting a more detailed data policy for a specific participation level.}
\begin{enumerate}[label=,leftmargin=*,labelsep=0mm]
    \item[ 
]\textbf{Policy Challenge: Agreements.} A single data policy is a major challenge. Data proprietary period conditions should be negotiated and~the term ``public'' data defined. ``Partial integration'' clients should integrate in a custom mode. 
This requires a balance between the Hub's need for standardisation and the client's need for flexibility.
    \item[ 
] \textbf{Enabling Synergistic Science.} An agreement is needed that unlocks the option for telescopes to work together as a single, more powerful instrument. 
    \item[ 
]\textbf{Fusion of Data.} An agreement is needed to confidently combine data from different instruments. 
\end{enumerate}

In Table~\ref{tab:assyhubfairprinciples} we summarise the FAIR criteria, highlighting the integration mechanisms to be used in the proposed participation levels of the~AstroHub.

There is a strategic need for FAIR principles to be implemented at every stage of AstroHub development. Thus, we are ensuring that the platform supports collaborative and data-driven astronomical~science.

\begin{table}[H]
    \footnotesize
    \caption{FAIR principles shaping the AstroHub participation~levels.}
    \label{tab:assyhubfairprinciples}
    \begin{adjustwidth}{-\extralength}{0cm}
        \begin{tabularx}{\fulllength}{m{2cm} m{5cm} m{5cm} m{5cm}<{\raggedright}}
        \toprule
            \textbf{Principles} 
 & \textbf{No Integration} & \textbf{Partial Integration }& \textbf{Full Integration} \\
        \midrule
            Findable 
      & Client's responsibility. AstroHub only knows the telescope exists.  & Shared data is indexed in the AstroHub catalogue. Private data is not.        & \textls[-20]{Mandatory and automated. All data is indexed in the central catalogue with a maximal set of metadata.}      \\
    \midrule
            Accessible    & Client's responsibility. Data is private, behind~the client's firewall. & Custom Access Control. The~AstroHub API enforces custom rules (e.g., project-based access and longer \mbox{proprietary periods).}  & Standardised Access Control. The~AstroHub API enforces the standard policy (e.g., 12-month proprietary, \mbox{then public).} \\
\midrule
            Interoperable & Client's responsibility. Any data format is accessible.  & Conditional. Data must meet AstroHub standards if it is to be processed by AstroHub pipelines or shared. An~option of a ``Validator Service'' to check new data.      & Mandatory and enforced. All data must pass through the \linebreak  standardising~pipeline. \\
\midrule
            Reusable      & Client's responsibility.  & Custom. Shared data must have a clear license and basic provenance.   & Mandatory and automated. The~pipeline generates detailed provenance. Standard licenses are automatically~applied. \\
        \bottomrule
        \end{tabularx}
    \end{adjustwidth}
\end{table}
\unskip

\section{Assy-Turgen AstroHub Development~Status}\label{sec:developmentstatus}
The performance of a ground-based telescope is inextricably linked to the quality of its location. One of such sites with a high potential is the Assy-Turgen Observatory\endnote{Assy-Turgen Observatory: \url{https://fai.kz/observatories/assy-turgen} (accessed on 15 July 2025).} of the Fesenkov Astrophysical Institute (FAI). In~this section, we will report on the status of the AstroHub case study, list currently operating telescopes, and~emphasise the integration strategy on how diverse instruments will be~united.

\subsection{Assy-Turgen Observatory Location, Area, and~Astroclimate}\label{sec:assylocationandarea}

\begin{figure}[H]
        
\begin{adjustwidth}{-\extralength}{0cm}
\centering 
\includegraphics[width=0.98\linewidth]{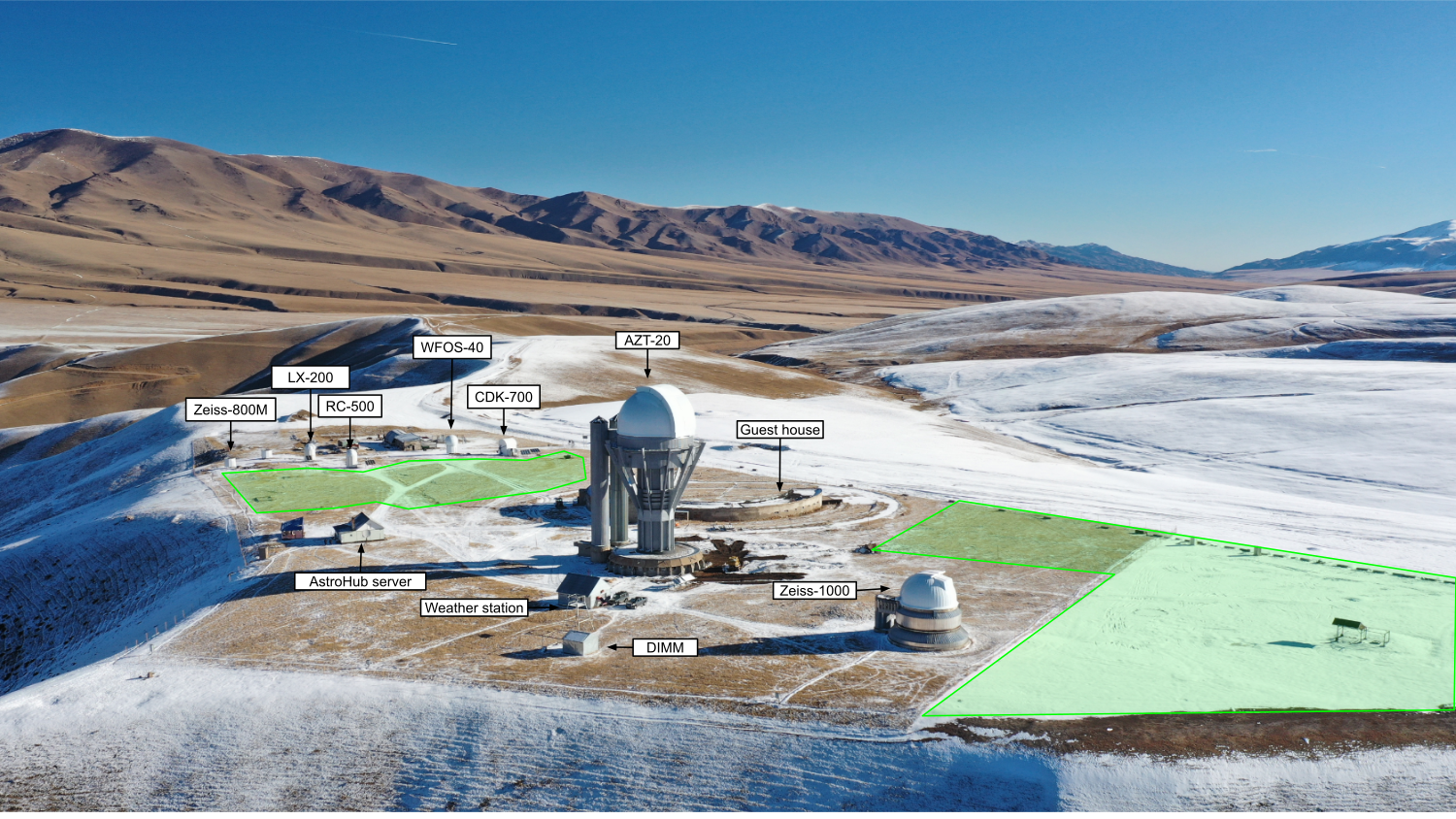}
\end{adjustwidth}

 \caption{Panoramic view of the Assy-Turgen Observatory with marked instruments by the end of 2024. Prospective sites for future medium- and small-aperture telescopes are highlighted in~green.\label{fig:Assy}}

\end{figure}

The Assy-Turgen Observatory (ATO) is located in the foothills of the Zailiyskiy Alatau, near~the watershed of the Assy and Turgen rivers, approximately $100$~km from Almaty. Its area is $\sim$2.3~hectares and can potentially host more than a hundred medium- and small-aperture telescopes. Horizon obstruction due to surrounding mountain ranges is minimal, averaging just $3.5$~degrees (Figure~\ref{fig:Assy}). Situated at an altitude of $2750$~m above sea level, the~observatory benefits from reduced atmospheric interference. On~average, there are $178$~clear nights per year, totalling $\sim$1500~clear night hours~annually.

\begin{figure}[H]
    \centering
    \begin{adjustwidth}{-\extralength}{0cm}
        \subfloat[\centering Wind speed]{\includegraphics[width=0.49\linewidth]{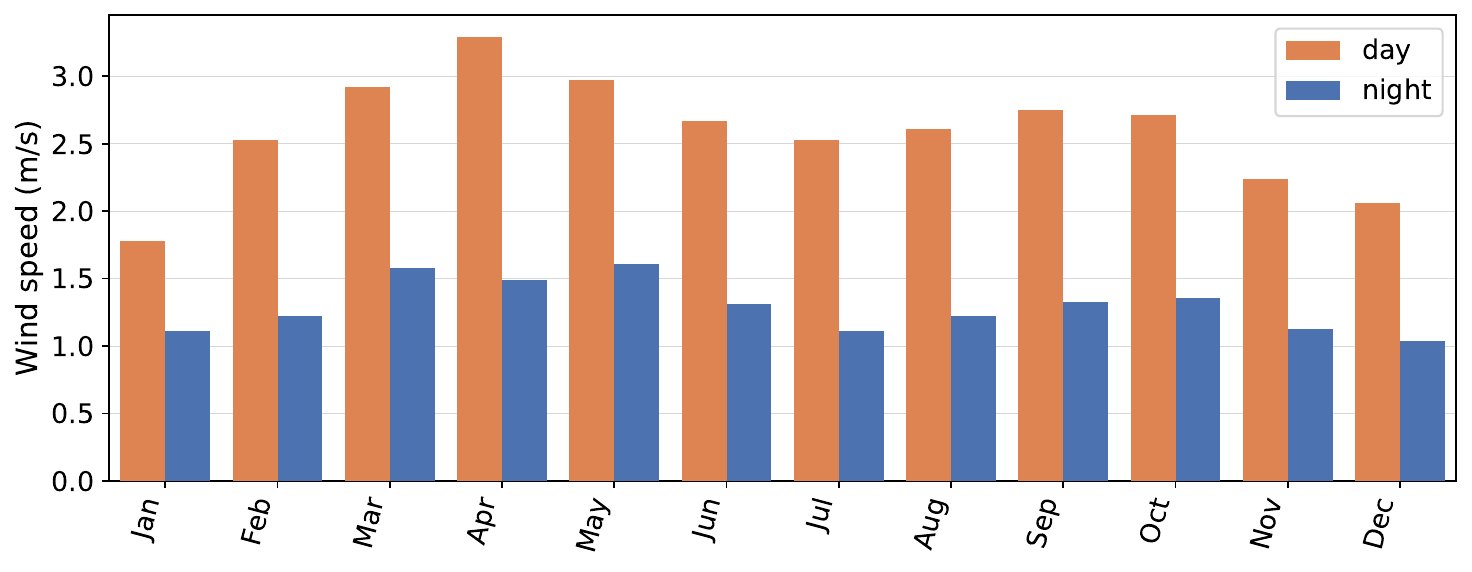}}
        \hfill
        \subfloat[\centering Humidity]{\includegraphics[width=0.49\linewidth]{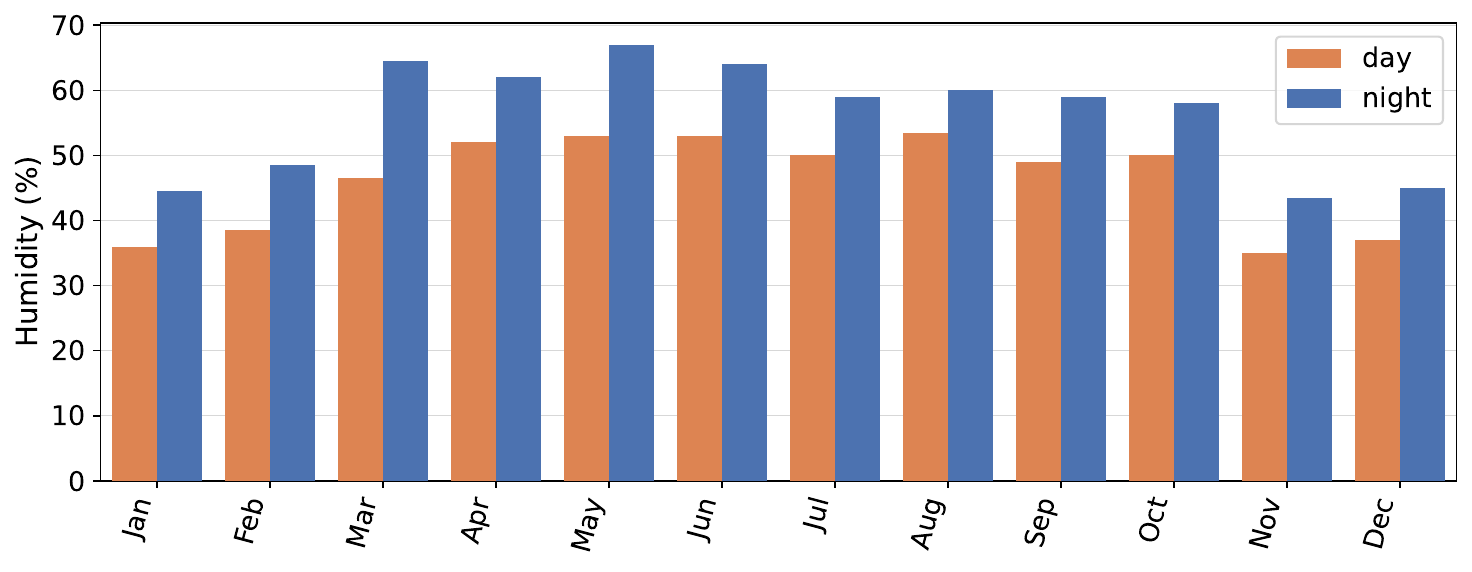}} 
         \end{adjustwidth}

\caption{\textls[-15]{Monthly median day/night values for wind speed (\textbf{a}) and humidity (\textbf{b}), measured at the Assy-Turgen Observatory during 2021--2025. Original data were measured with a 2-s~resolution.}}

        \label{fig:monthlystats}
   
\end{figure}

\begin{figure}[H]
       
\begin{adjustwidth}{-\extralength}{0cm}
\centering 
 \includegraphics[width=0.9\linewidth]{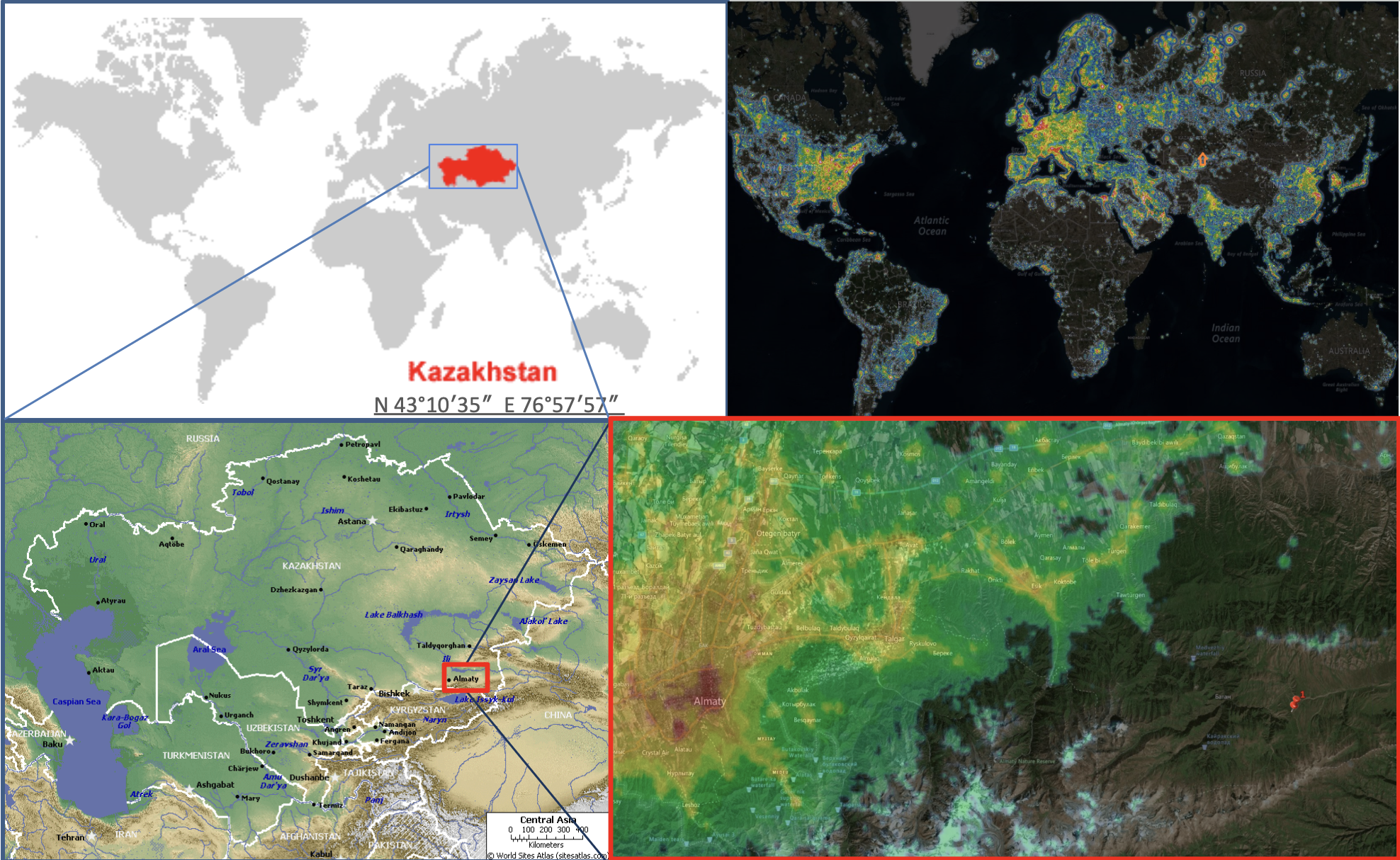}
\end{adjustwidth}

\caption{Light pollution 
 map of the world (upper right panel) and near-Almaty region (lower right panel). The~Assy-Turgen Observatory is marked with a red pin {on the lower right panel. Data were retrieved from the light pollution map.\protect\endnote{Light-pollution map: \url{https://www.lightpollutionmap.info} (accessed on 10 June 2025).}}}

        \label{fig:lightpollution}
   
\end{figure}

The local climate is characterised by an average nighttime air temperature of +9.3\,$^\circ$C in summer and $-$10.5\,$^\circ$C in winter, with~an average diurnal temperature variation of $5\,^\circ$C. The~average wind speed on clear nights is $1.7$~m/s at a height of $10$~m, see Figure~\ref{fig:monthlystats}.  The~mean night sky brightness in the $3900$--$6000\,\AA$~range is 22\,mag/arcsec$^2$, measured in zenith utilising the Sky Quality Meter (SQM-L)\endnote{Sky Quality Meter: \url{https://www.unihedron.com/projects/darksky} (accessed on 15 July 2025).} \cite{SanchezdeMiguel...2017MNRAS.467.2966S}. For~comparison, the~estimates that were obtained in the Almaty urban region do not exceed 18 mag/arcsec$^2$. The~absence of dust storms, the~relative stability of the atmospheric layer, and~the exceptionally dark night sky make the Assy-Turgen Observatory a favourable site for high-quality astronomical observations, see Figure~\ref{fig:lightpollution}.

{We used a Different Image Motion Monitor (DIMM) in the test regime to determine a seeing parameter on the Assy-Turgen plateau. Technical characteristics of the DIMM are described in Section~\ref{sec:assyinstruments}. 
 For~seeing calculations, we used a standard method and procedure described in \citep{Tokovinin...2007MNRAS.381.1179T}. We alternated between taking $200$~snapshots with a $2$~ms and $200$~snapshots with a $4$~ms exposure time for each of five randomly selected bright stars with a zenith distance not exceeding $30$~degrees. Our test measurements yielded a median seeing value of $1.5\pm0.3\,\text{arcsec}$.}

\vspace{-15pt}

\begin{figure}[H]
        
\begin{adjustwidth}{-\extralength}{0cm}
\centering 
\includegraphics[width=0.9\linewidth]{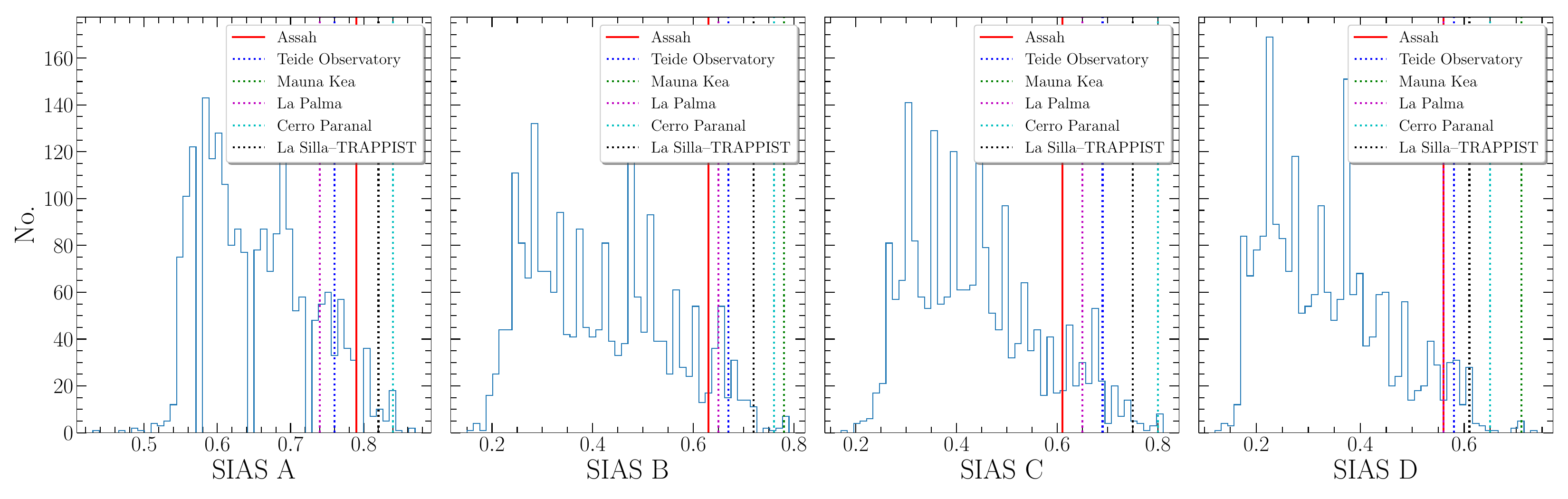}
\end{adjustwidth}

\caption{The distribution of the Suitability Index for Astronomical Sites (SIAS) parameters across 2122 observatories worldwide, according to the SIAS catalogue~\cite{Aksaker...2020MNRAS.493.1204A}. Parameters for the Assy-Turgen Observatory are marked by a vertical solid red line, while coloured dotted lines indicate other \mbox{leading observatories.}}

        \label{fig:assysiasworld}

\end{figure}

Modern observatory site selection employs a formal process known as Multi-Criteria Decision Analysis (MCDA) \citep{ishizaka2013multi}, often in conjunction with Geographic Information Systems (GISs), to~systematically evaluate and rank potential locations. This framework enables the integration and weighting of numerous, often conflicting, criteria. We utilised a site selection database described in~\cite{Aksaker...2020MNRAS.493.1204A}, with the distribution of the Suitability Index for Astronomical Sites (SIAS) parameters across different observatories worldwide. The~SIAS parameters for the Assy-Turgen Observatory (designated in the database as ``Assah'', MPC code 217) are indicated by the red dashed line in Figure~\ref{fig:assysiasworld}. For~comparative reasons, we also mark a few world-leading~observatories.

\vspace{-9pt}

Numerical values of the SIAS parameters are also given in Table~\ref{tab:assysiasparameters}. One can clearly see that across observatories worldwide, the~ATO, placed in the leading quartile by all four SIAS criteria, positions itself as a very good site for astronomical~observations.

\begin{table}[H]
    \small
    \caption{Suitability Index for Astronomical Sites (SIAS) values and its constituent parameters (layers) for the Assy-Turgen Observatory (denoted ``Assah'', MPC code 217), according to the catalogue~\cite{Aksaker...2020MNRAS.493.1204A}. Other Kazakhstani observatories are also listed for comparison purposes: TSHAO---Tien-Shan Astronomical Observatory; Alma-Ata---Kamenskoye Plateau~Observatory.}
    \label{tab:assysiasparameters}
    
    \scriptsize
    \begin{adjustwidth}{-\extralength}{0cm}
        \begin{tabularx}{\fulllength}{LLlLLLlLlLLLLl}
        \toprule
            \vspace{-6pt}\textbf{Long. deg} & \textbf{Lat. deg} & \textbf{CC} \boldmath\textbf{\%} & \textbf{DEM} \textbf{m} & \textbf{AL}  \textbf{W cm}\boldmath{$^{-2}$} \textbf{sr}\boldmath{$^{-1}$} & \textbf{PWV} \textbf{mm} & \textbf{AOD} & \textbf{WS} \boldmath\textbf{m\,s$^{-1}$} & \textbf{LULC} & \textbf{SIAS A} & \textbf{SIAS B} & \textbf{SIAS C} & \textbf{SIAS D} & \textbf{Observ.}\\

            \midrule
            77.87 & 43.23 & 0.42 & 2662.00 & 5.72 & 0.00 & 0.12 & 2.54 & 10 & 0.79 & 0.63 & 0.61 & 0.56 & Assah\\
            76.97 & 43.06 & 0.40 & 2581.00 & 4.26 & 0.45 & 0.20 & 1.73 & 10 & 0.78 & 0.61 & 0.60 & 0.54 & TSHAO\\
            76.96 & 43.19 & 0.47 & 1189.00 & 6.01 & 5.05 & 0.25 & 0.86 & 10 & 0.66 & 0.37 & 0.41 & 0.32 & Alma-Ata\\  
        \bottomrule
        \end{tabularx}
   
\begin{adjustwidth}{+\extralength}{0cm}
 \footnotesize{
        Abbreviations and minimum--maximum values for the entire catalogue: Long.---longitude; Lat.---latitude; \linebreak  CC---cloud coverage (min: 0.3, max: 0.95); DEM---digital elevation model (min: $-408$, 
 max: 8685); \mbox{AL---artificial} light (min: 0, max: 1682.71); PWV---precipitable water vapour (min: 0.68, max: 53.54); AOD---aerosol optical depth (min: 0, max: 4.98); WS---wind speed (min: 0.01, max: 84.67); LULC---land use and land cover; \linebreak  Observ.---observatory.
    }
\end{adjustwidth}
    \end{adjustwidth}
\end{table}
\unskip

\subsection{Prerequisites for AstroHub at~ATO}\label{sec:assyprerequisites}
Over the past six years, the~number of instruments installed at ATO has increased from three to seven, with~plans to double the current park of telescopes in the next three years, see Table~\ref{tab:assyinstrumentsinfo}. The~work on the integration of these tools and, most importantly, the~relevant research groups into a single interconnected ecosystem is a step that should be performed after two prerequisite~projects:
\begin{enumerate}
    \item ``Development of a national Virtual Observatory based on robotic telescopes, Big Data technologies and high-performance computing systems'' (2021--2023);\endnote{Kazakhstani Virtual Observatory: \url{https://fai.kz/projects/virtobs} (accessed on 15 July 2025).} 
    \item ``Development of Kazakhstani telescope network for the national space situational awareness system'' (2023--2025).\endnote{Telescopes Network project: \url{https://fai.kz/projects/telnet} (accessed on 15 July 2025).}
\end{enumerate}
The idea behind the first project (KazVO, for~short) was to advance astronomical research to a new technological level, integrating instrumental capabilities and observational data into the international astronomical environment through the development of a national Virtual Observatory as part of the IVOA.\endnote{International Virtual Observatory Alliance: \url{https://www.ivoa.net} (accessed on 15 July 2025).} The latter project (Telescopes Network, for~short) is intended to develop a multifunctional optical observatory with flexible scheduling, designed for research in near-Earth space. As~a result, favourable conditions were developed at the~ATO.

\subsection{Instruments}\label{sec:assyinstruments}
\begin{figure}[H]
    \centering
    \begin{adjustwidth}{-\extralength}{0cm}
        \subfloat[\centering]{\includegraphics[width=7cm]{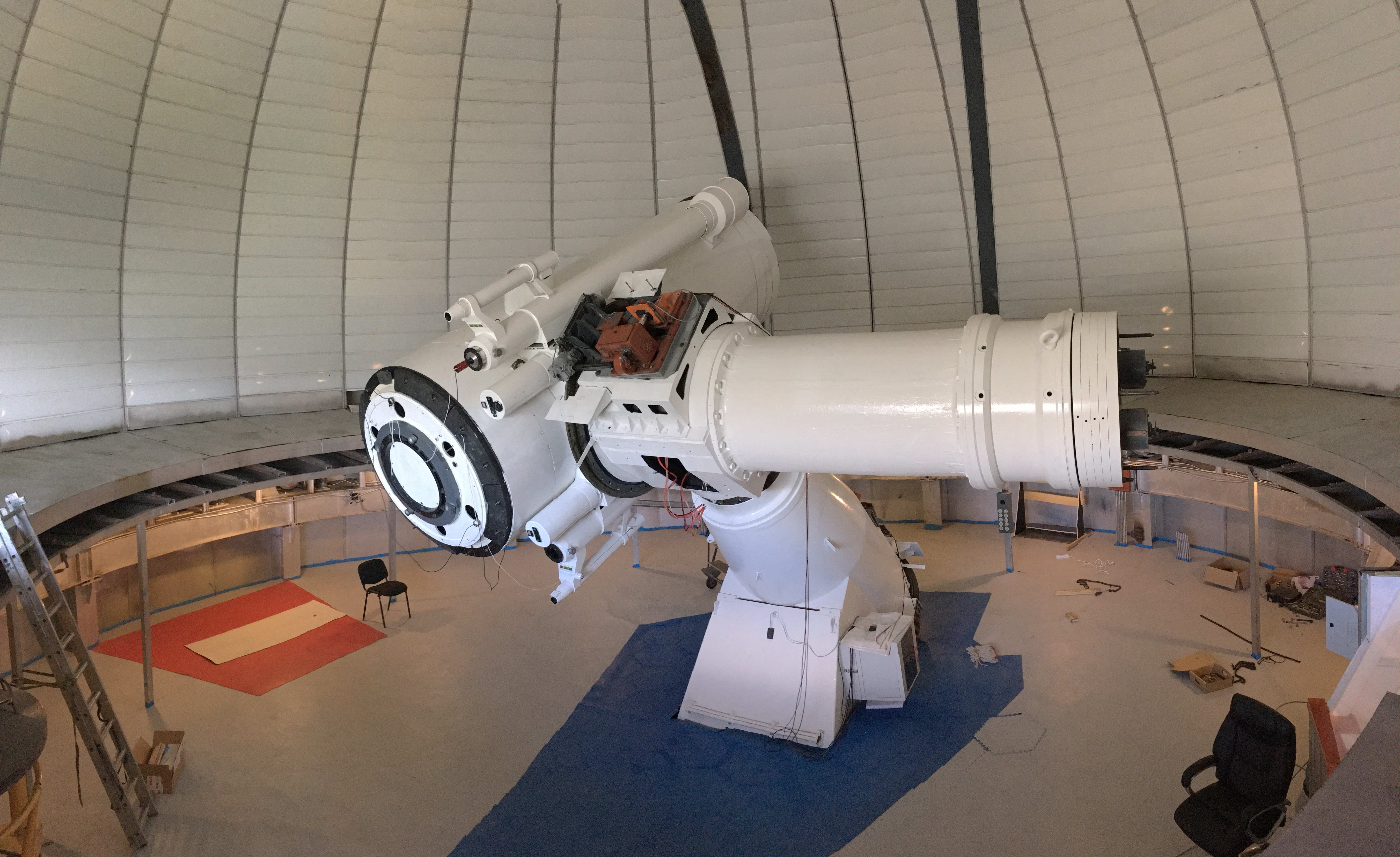}}
        \hfill
        \subfloat[\centering]{\includegraphics[width=6cm]{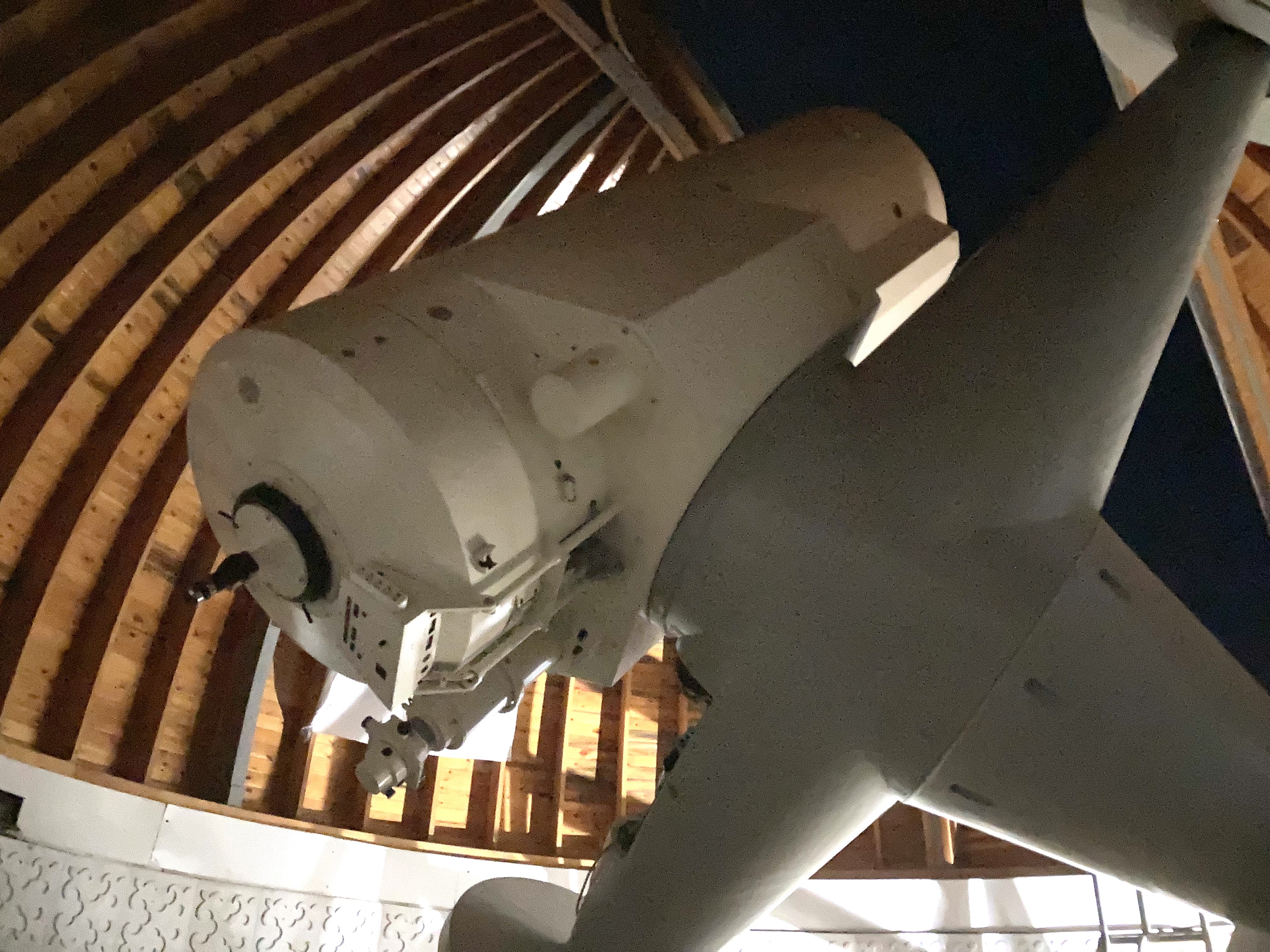}} 
        \hfill
        \subfloat[\centering]{\includegraphics[width=5cm]{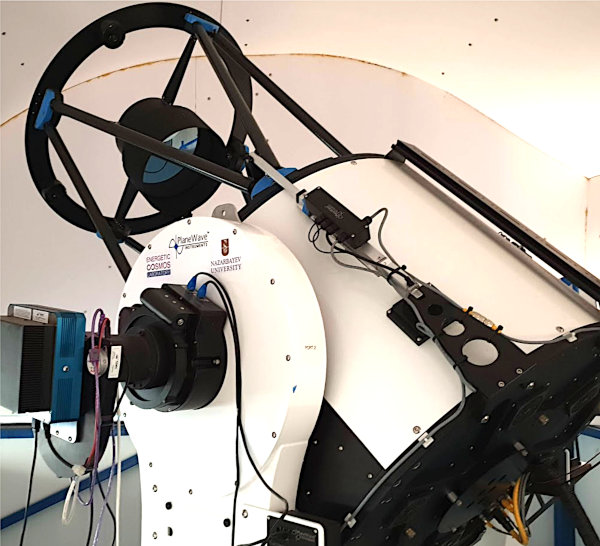}}\\
        \subfloat[\centering]{\includegraphics[width=5cm]{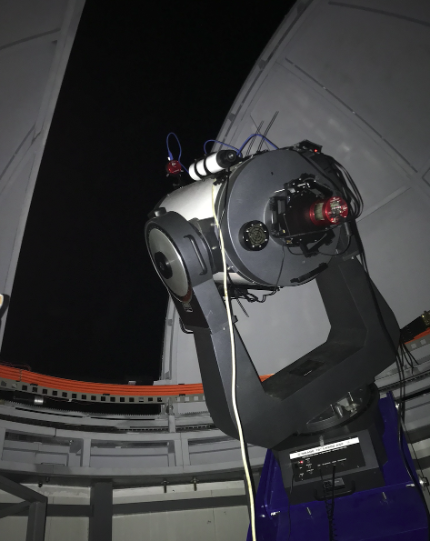}}
        \hfill
        \subfloat[\centering]{\includegraphics[width=6.5cm]{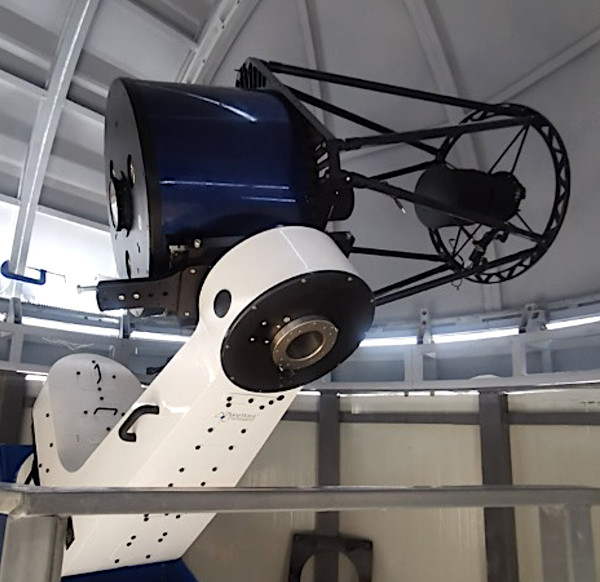}}
        \hfill
        \subfloat[\centering]{\includegraphics[width=5.15cm]{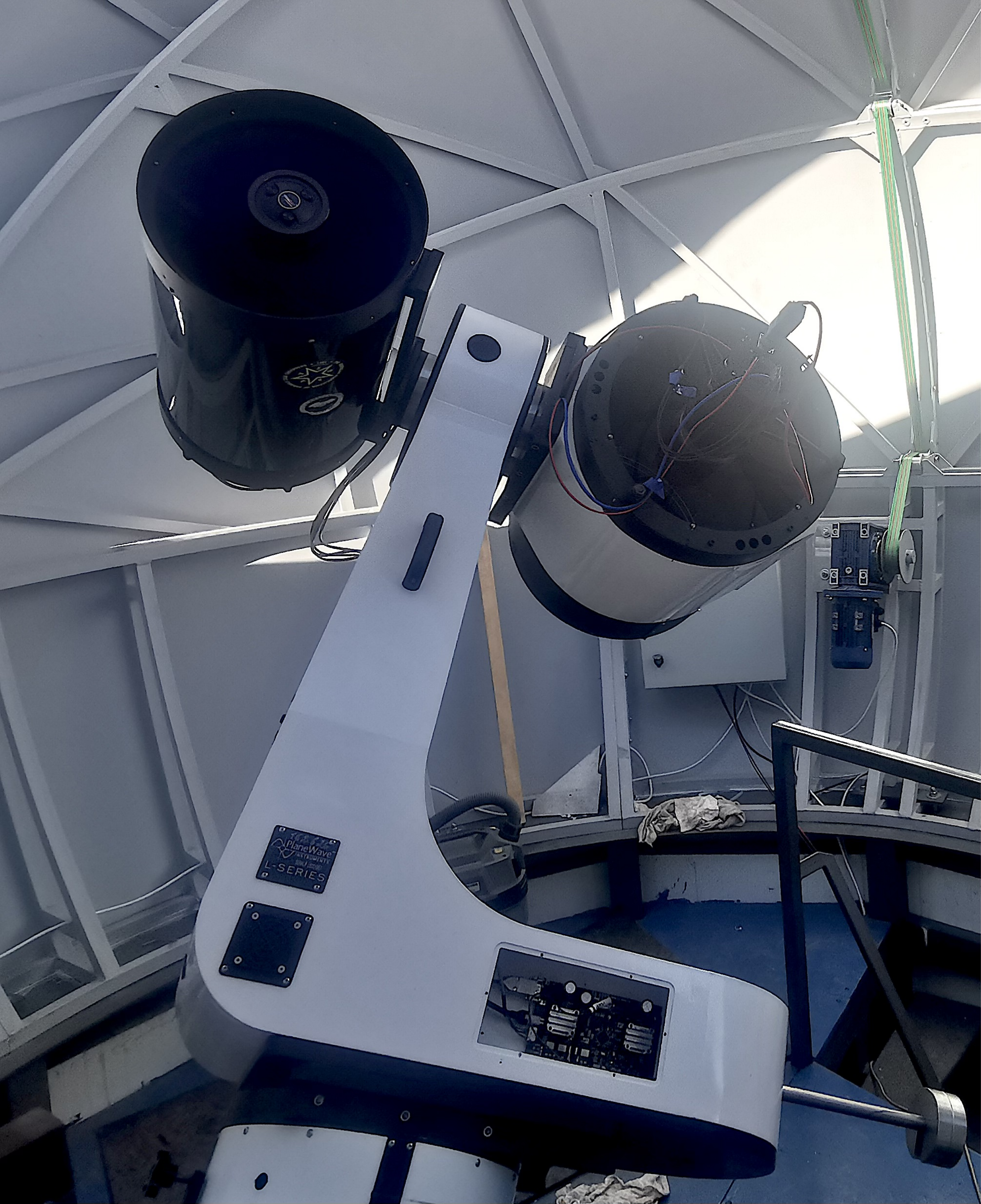}}

    \end{adjustwidth}
        \caption{Currently operated instruments at the ATO: (\textbf{a})~AZT-20; (\textbf{b})~Zeiss-1000; (\textbf{c})~CDK-700; (\textbf{d})~LX-200;  (\textbf{e})~RC-500; and (\textbf{f})~WFOS-40. \label{fig:assyinstrumentspictures}}

\end{figure}

Currently, there are seven optical telescopes up to 1.5 m aperture, all equipped with professional CCD cameras, located on ATO, see Table~\ref{tab:assyinstrumentsinfo} and Figure~\ref{fig:assyinstrumentspictures}. Two telescopes are mounted on PlaneWave L500 and PlaneWave L600 Direct Drive and dedicated to NEOs and LEOs. It is planned that currently running and future telescopes will be given specific roles for better interoperability and coordination within the \textit{full integration} telescope~network.

The Assy-Turgen Observatory also has an autonomous weather monitoring system, Sky Alert, for~automated observatories and~a professional weather station, Davis Vantage Pro2. Information from these devices provides high-accuracy data on the meteorological situation at the observatory (see Figure~\ref{fig:monthlystats}) and the level of atmospheric~transparency.

\begin{table}[H]
 \scriptsize
    \caption{Characteristics of the currently operating optical telescopes and~instruments.}
    \label{tab:assyinstrumentsinfo}
    \begin{adjustwidth}{-\extralength}{0cm}
        \begin{tabularx}{\fulllength}{llllLLLLl}
        \toprule
        \textbf{Telescope} & \textbf{D} & \textbf{f} & \textbf{Obs.}  & \textbf{Gratings/Filters} & \textbf{Detector} & \textbf{FoV/Slit }& \textbf{Obs.} & \textbf{Client \textsuperscript{2}/} \\
        
        & \textbf{[mm]} & \textbf{[mm]} & \textbf{Type \textsuperscript{1}} & \textbf{[lines/mm]} & & \boldmath\textbf{[$^\circ/'/''$]} & \textbf{Mode} & \textbf{Commis. \textsuperscript{3}} \\
        \midrule
            AZT-20     & $1560$ & $5720$   & Spec/Phot & $360/1800/2400$ g'r'i'z' & EMCCD~Andor/CMOS Kepler KL400 & $3''/5''/9''$ $14'\times14'$ & Follow-up/Program  & FAI/2017 \\
            Zeiss-1000 \textsuperscript{4} & $1016$ & 13,300  & Polar     & ---                      & ---                     & ---                        & Follow-up/ Program  & FAI/2027 \\
            Zeiss-800M & $800$  & $2096$   & Phot      & g'r'i'                   & CMOS QHY-6060 Pro       & $1.3^\circ\times1.0^\circ$ & Follow-up/Program  & FAI/2024 \\
            CDK-700    & $700$  & $4540$   & Phot      & g'r'i'                   & $3\times$EMCCD N\"uv\"u & $10'\times10'$             & Alert/Follow-up    & NU/2019 \\
            WFOS-70    & $660$  & $900$    & Phot      & clear                    & CMOS QHY-6060 Pro       & $3.8^\circ\times3.8^\circ$ & Follow-up          & FAI/2025 \\
            RC-500     & $508$  & $1400$   & Phot      & clear                    & CMOS QHY-600            & $1.3^\circ\times1.0^\circ$ & Alert/Follow-up    & FAI/2019 \\
            LX-200     & $406$  & $4064$   & Phot      & BVR, H$\alpha$, OIII     & CCD ATIK 16200          & $24' \times 19'$           & Program/Astrometry & Pulkovo/2022\\
            WFOS-40    & $400$  & $550$    & Phot      & clear                    & CMOS QHY-600            & $3.8^\circ\times2.5^\circ$ & Survey             & FAI/2024 \\
            DIMM \textsuperscript{5}      & $280$  & $2240$   & Phot      & clear                    & CMOS Point Grey High-Speed & ---                     & Program            & FAI/2024 \\
        \bottomrule
        \end{tabularx}
    
\begin{adjustwidth}{+\extralength}{0cm}
\footnotesize{
        \textsuperscript{1} Observation 
 Type: Spec---spectroscopy; Phot---photometry; Polar---spectropolarimetry.         \textsuperscript{2} Client and Instrument Owner: FAI---Fesenkov Astrophysical Institute (Almaty, Kazakhstan); NU---Nazarbayev University (Astana, Kazakhstan); Pulkovo---Main Pulkovo Astronomical Observatory (Saint Petersburg, Russia).         \textsuperscript{3} Year of commissioning of the instrument. \textsuperscript{4} Zeiss-1000: Spectropolarimeter instrument is under development. \textsuperscript{5} DIMM: Currently at testing stage, {with measurements of seeing of $1.5\pm0.3\,\text{arcsec}$ (August 2025)}}.
\end{adjustwidth}
    \end{adjustwidth}
\end{table}

\vspace{-18pt}

\unskip 

\subsubsection*{Computing and Data Storage~Cluster}\label{sec:astrohubhpc}
\begin{figure}[H]
    \includegraphics[width=0.7\textwidth]{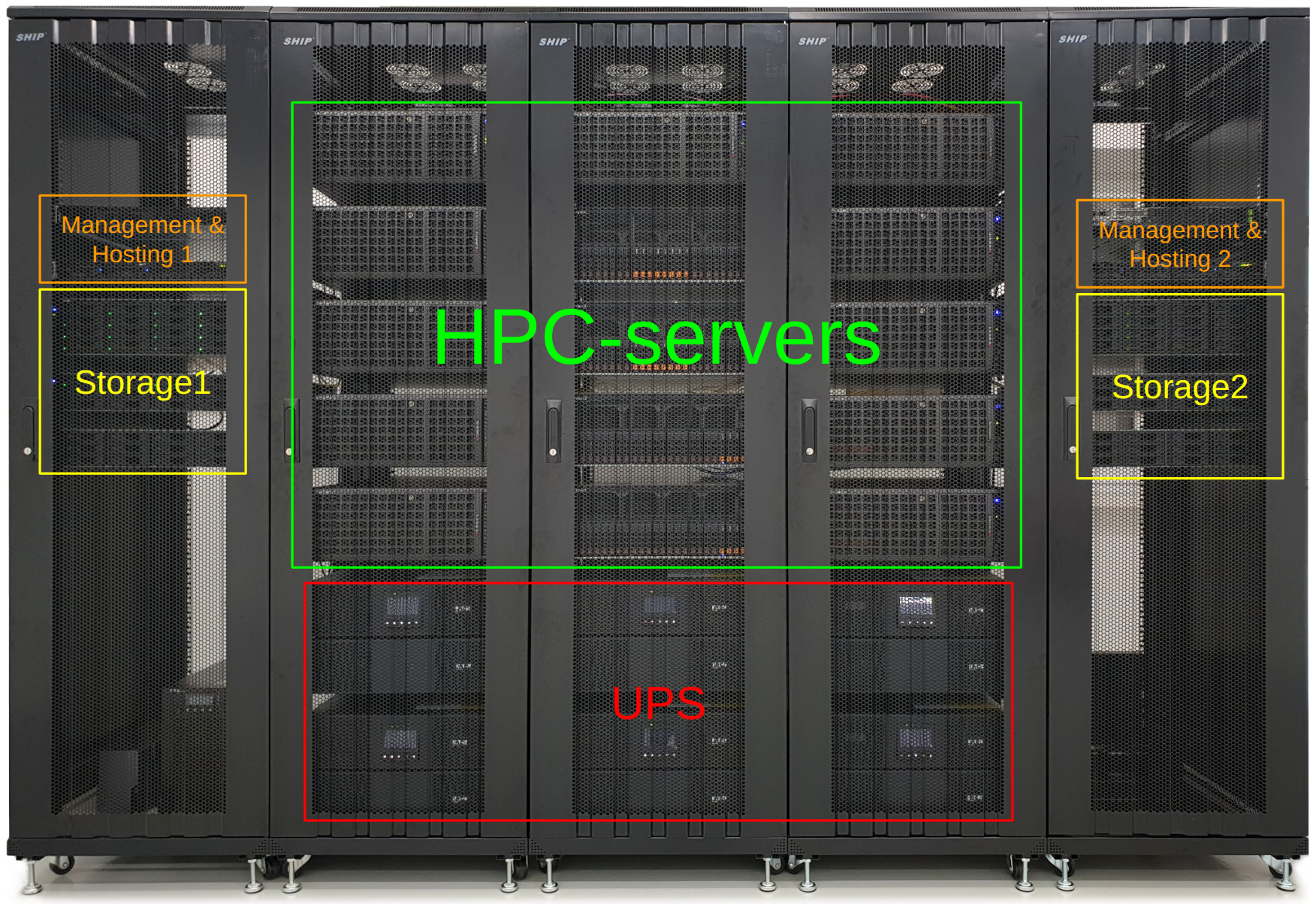}
    \caption{FAI's computer~cluster.}
    \label{fig:fai-cluster}
\end{figure}

FAI operates a computer cluster (Figure~\ref{fig:fai-cluster}) comprising several high-performance computing nodes equipped with high-end multi-core CPUs and GPU cards.\endnote{FAI's computer cluster: \url{https://fai.kz/instruments/computer-cluster} (accessed on 15 July 2025).} The theoretical single-precision performance of the cluster currently peaks at $127.2$~TFLOPS for CPU operations ($1268$~cores/$2536$~threads) and $2287$~TFLOPS for GPU operations (501,248~CUDA cores). In~addition, it has a redundant data storage of $270$~Terabytes in volume. Cluster has a 10-Gigabit interlink and operates under a Linux-based OS. Task scheduling is organised using SLURM (Simple Linux Utility for Resource Management) workload manager, and~inter-host parallel computations are managed by~an MPI.

The cluster plays a pivotal role in the AstroHub project: it provides the hosting for all related services, the~space for the long-term data storage, and~computational resources for data processing. For~external users, computation and data storage operations on the FAI cluster can be arranged via online task submission that, once approved, results in an SSH account having access to the master node with agreed~permissions. 

\begin{figure}[H]
    \includegraphics[width=0.8\textwidth]{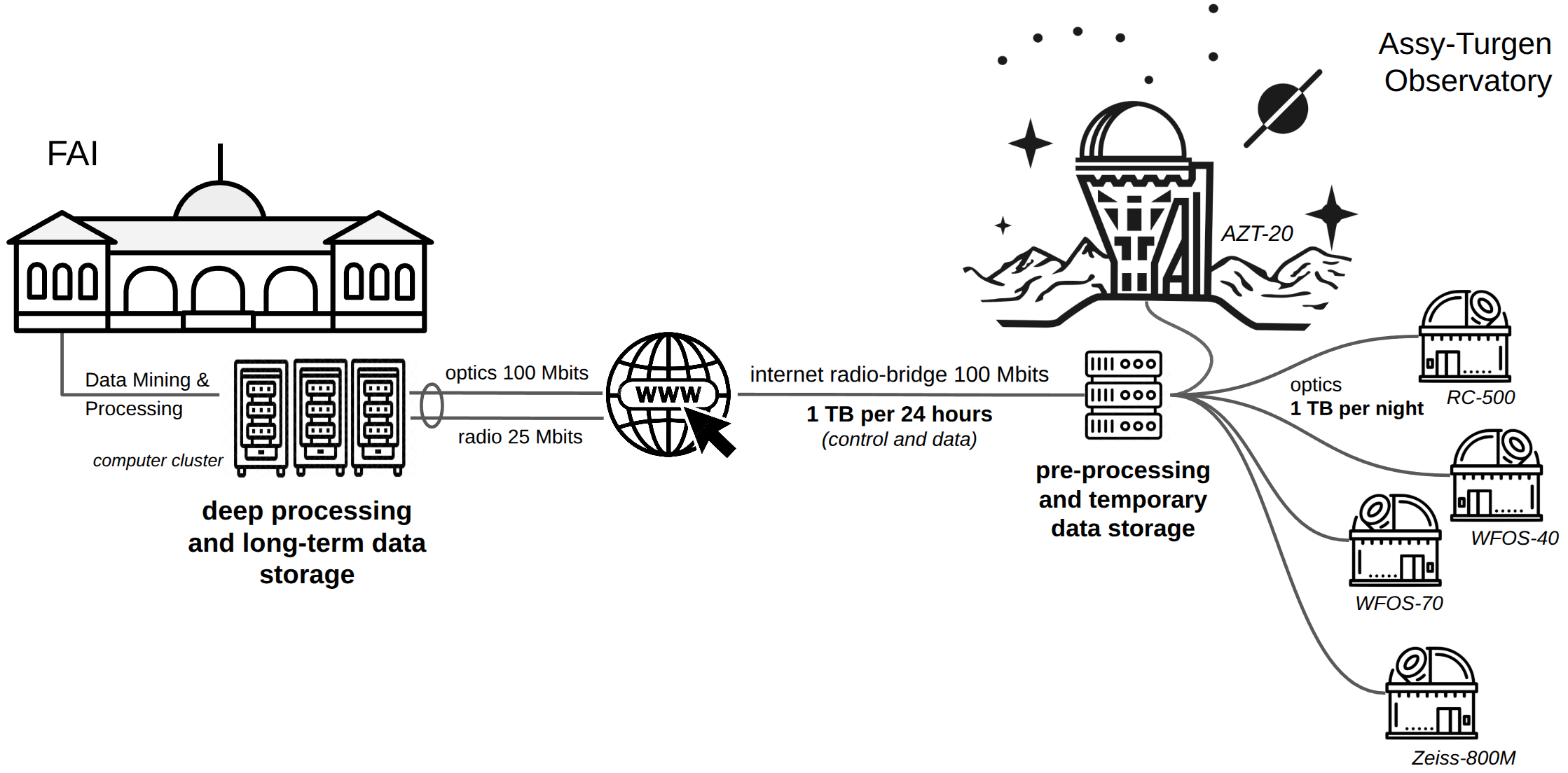}
    \caption{FAI-ATO connection scheme. FAI hosts and operates the main HPC with dedicated power and storage for AstroHub. 
    The Assy-Turgen Observatory has an on-site server for observatory platform operation (scheduling, telescope operation, data transfer, etc.), temporary data storage, \mbox{and processing.}}
    \label{fig:fai-oat-relation}
\end{figure}

Internet access for the FAI cluster is provided by two independent Internet connections (Figure~\ref{fig:fai-oat-relation}): the main and the backup one. The~main connection is a fiber-optic line with $100$~Mbps bandwidth. The~backup connection is a $25$~Mbps radio-bridge line. Both channels operate in load balancing mode, providing fault tolerance and increasing the effective bandwidth up to $125$~Mbps. If~one of the channels fails, all connections passing through it are automatically re-established through the second channel, ensuring redundancy of the communication line with the~cluster. 

The cluster, started about a decade ago from a small computing server blocks~\cite{Bibossinov...2018CoBAO..65..124B}, has now become a standalone computational unit, which supports all current research projects in FAI, including AstroHub. Its modular architecture allows one to increase the number of computing servers and GPU cards and the storage capacity and~interlink speed whenever possible. Having a server unit at the ATO for local AstroHub routines (telescopes network operation, short-term data storage, and~processing, server tasks, etc.) and the cluster at FAI's main building for long-term data storage and deep data analysis all enable AstroHub science tasks to be solved in the most convenient and secure~way.

\subsection{AstroHub Interoperability at~ATO}\label{sec:assyinteroperability}
The development of a general scheme of interoperability within the AstroHub at the Assy-Turgen Observatory is focused on the efficient use of resources and cooperation between the operator (FAI) and clients, the~AstroHub users. Work in this direction entails the parallel development of interaction between hardware and software, as~well as the development and consolidation of official documents and agreements that regulate participation in events based on AstroHub, as~discussed in Section~\ref{sec:fairandvo}.

\begin{figure}[H]
    \includegraphics[width=0.7\textwidth]{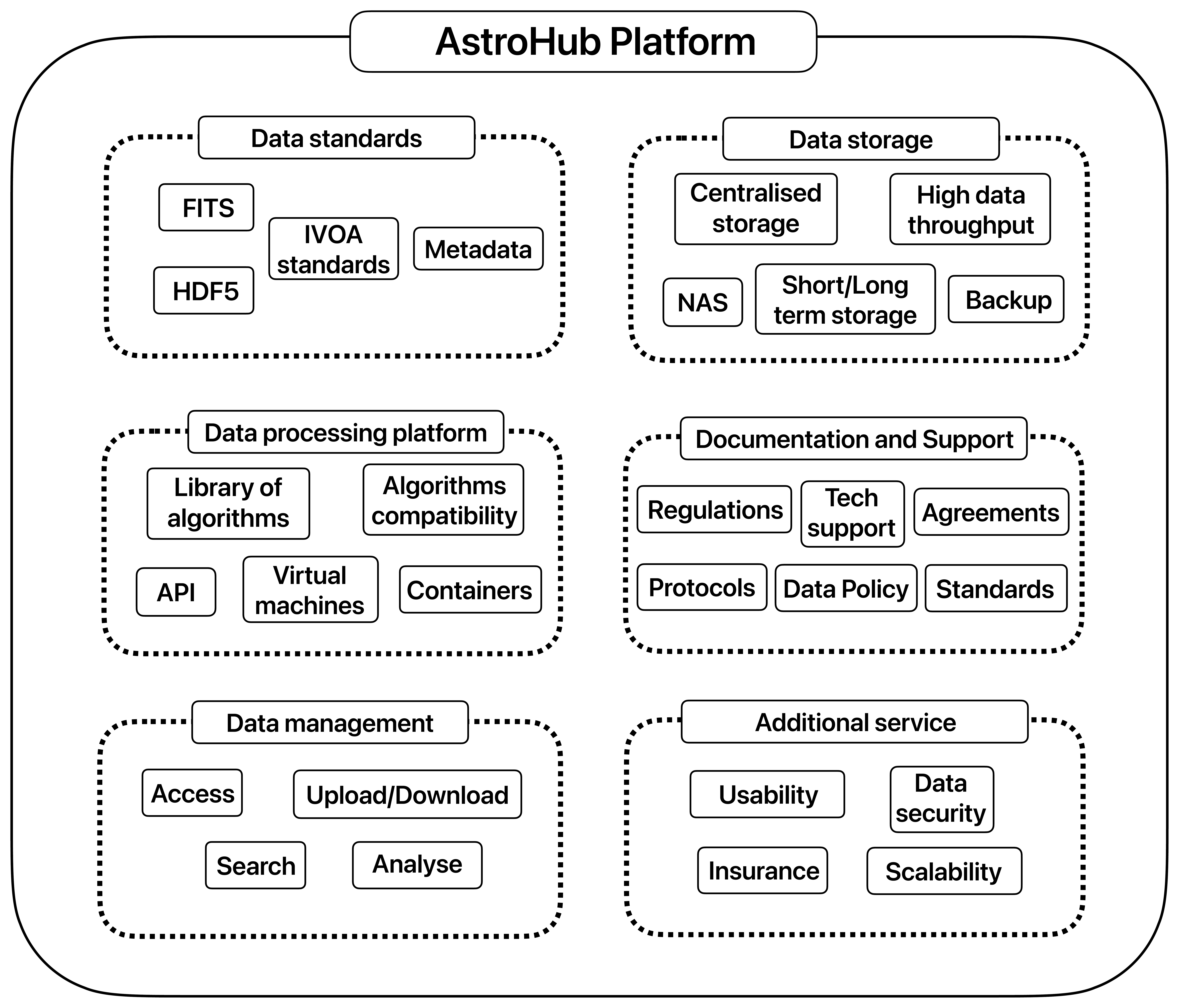}
    \caption{AstroHub Platform development scheme at~ATO.}
    \label{fig:astrohub_platform}
\end{figure}

It is assumed that FAI, as~the operator, will provide a platform for clients---research individuals, groups, and organisations---interested in installing their instruments at the ATO, as~well as a service for monitoring observations and obtaining high-quality scientific data. We have developed an interoperability scheme, which includes the following main solutions, see Figure~\ref{fig:astrohub_platform}:

\begin{enumerate}
    \item  Standardisation of data formats. Adoption of a single format for storing observational data: At this stage, it was decided to continue to use the FITS (Flexible Image Transport System) format, which remains the most common standard in astronomy. At~the same time, the~HDF5 format is also promising, especially for large volumes of data. The~main criterion for the data format was defined as support for metadata describing the observation~conditions.
    
    Development of a metadata standard will ensure convenient searching, filtering, and comparison of data from different telescopes and instruments. At~this stage, it was decided to use and develop the existing IVOA standards, given the international scale of AstroHub~operations.

    \item Centralised data storage. The~AstroHub operator will provide clients with centralised storage with high bandwidth and reliable backup. This will ensure the safety of data and facilitate access to it for all clients. The~existing infrastructure (fiber-optic cables) already provides high bandwidth for data transmission from the observatory instruments. At~this stage, it was decided to deploy a centralised storage with a backup~system.
    
    Development of a data management system. The~centralised system should allow clients to upload, store, search, access, and process data. Following the integration levels, access and rights will be differentiated and data sharing will be~possible.
    
    \item Data Processing Platform: By design, the~local platform will provide access to a wide range of tools for processing astronomical data, including calibration, reduction, analysis, and visualisation (see also Appendix~\ref{sec:patents}).
    \begin{itemize}
        \item Development of a library of algorithms: The operator on the platform provides the opportunity for clients to share their processing algorithms with other participants, which facilitates collaboration and accelerates scientific research.
        \item Development of mechanisms for calling and using algorithms: This solution can be implemented through an API (Application Programming Interface) or other integration tools. 
        \item Ensuring compatibility of algorithms with the data format and processing platform: The developed algorithms will be integrated into the platform with subsequent debugging for use on the central server, with~subsequent output of the task file and observation commands to each telescope.
        \item Use of virtual machines or containers (Docker): This solution will allow clients to run their algorithms in an isolated environment with the necessary dependencies, ensuring portability and reproducibility of results.
    \end{itemize}
    \item Documentation and support: The main regulations, standards, and protocols for the operator (FAI) to develop within the framework of information and communication interaction have been defined, including (but not limited to) the following:
    \begin{itemize}
        \item Creation of detailed documentation on data formats, the API, the processing platform, and other aspects of interoperability.
        \item Provision of technical support to clients for issues of integrating existing and new tools and using the system.
        \item Repair and routine maintenance of equipment and tools; provision of additional memory service in storage, expansion of the communication channel, data processing service, etc.
    \end{itemize}
    
    Documents under development for the current period include the following: \mbox{(1) protocol} for ordering and distributing observation time; (2) protocols for storing, moving, and publishing data obtained by AstroHub clients; (3) protocol for access to primary and reduced observational data; (4) data provision policy; (5) privacy policy and protection of personal information of AstroHub clients; and (6) policy of interaction between clients and the AstroHub~operator.
    
    \item Additional aspects~include the following:
    \begin{itemize}
        \item Data security. An~important aspect is to ensure the security and privacy of customer data.
        \item Scalability. The~system must be scalable to cope with the growing volume of data and number of customers.
        \item Usability. Interfaces must be intuitive and easy to use.
    \end{itemize}
\end{enumerate}

\unskip 

\section{Joint Research and International~Campaigns}\label{sec:researchandcampaigns}
The idea of hub unification is already pressing for the ATO, as~today the observatory is actively collaborating with institutes and universities worldwide, providing technological capabilities and a platform. In~this section, we outline the currently running observing programs and campaigns, as~well as our active research~connections. 

Table~\ref{tab:assyhubclientsplans} presents the current status of AstroHub client participation information. In~addition to those mentioned above, \textit{full}, 
 \textit{partial}, and \textit{no integration} levels, there is a \textit{decision} stage where the newcomers can set up the most convenient environment for \mbox{their instruments.}

\begin{table}[H]
    \footnotesize
    \caption{Assy-Turgen AstroHub client participation~information.}
    \label{tab:assyhubclientsplans}
    \begin{adjustwidth}{-\extralength}{0cm}
        \begin{tabularx}{\fulllength}{m{1.1cm} m{1.5cm} m{1cm} m{1.5cm} m{2.7cm} m{3.8cm} m{2.2cm} m{3cm}}
        \toprule
            \textbf{Client 
 \textsuperscript{1}} & \textbf{Country} & \textbf{Level \textsuperscript{2}} & \textbf{Instruments} &  \textbf{Observation \linebreak  Type \textsuperscript{3}} & \textbf{Observation Mode} & \textbf{Target \textsuperscript{4}} & \textbf{Commis. \textsuperscript{5}}\\
        \midrule
            Ariane      & France     & No        & $1$      & Phot            & Follow-up/Program       & NEO           & 2026\\
            Digantara   & India      & Decision  & Decision & Phot            & Survey                  & Decision      & 202x\\
            FAI         & Kazakhstan & Full      & $7$      & Phot/Spec/Polar & Alert/Follow-up/Program & Deep/NEO/LEO  & 2025\\
            China       & China      & Partial   & $>$1     & Phot            & Survey                  & Decision      & 202x\\
            KazNU       & Kazakhstan & Decision  & Decision & Phot            & Program                 & Deep          & 202x\\
            NU          & Kazakhstan & Partial   & $1$      & Phot            & Alert/Follow-up         & Deep          & 2019\\
            POLSA       & Poland     & No        & $1$      & Phot            & Survey                  & Decision      & 202x\\
            Pulkovo     & Russia     & Partial   & $1$      & Phot            & Program/Astrometry      & Deep/NEO      & 2022\\
Sybilla     & Poland     & Decision  & Decision & Phot            & Survey                  & NEO           & 202x\\

\bottomrule
\end{tabularx} 

\begin{adjustwidth}{+\extralength}{0cm}
\footnotesize{
        \textsuperscript{1} Client 
 and Instrument Owner: Ariane---ArianeGroup SAS (Les Mureaux, France
); Digantara---Digantara Research and Technologies (Bengaluru, India); FAI---Fesenkov Astrophysical Institute (Almaty, Kazakhstan); KazNU---Al-Farabi Kazakh National University (Almaty, Kazakhstan); NU---Nazarbayev University (Astana, Kazakhstan); POLSA---Polish Space Agency (Gda\'nsk, Poland); Pulkovo---Main Pulkovo Astronomical Observatory (Saint-Petersburg, Russia); Sybilla---Sybilla Tech (Poznań, Poland). \textsuperscript{2} AstroHub Participation Level (see details in the text): Full---full integration; Partial---partial integration; No---no integration; \mbox{Decision---deciding. 
\textsuperscript{3} Observation} Type: Spec---spectroscopy; Phot---photometry; \mbox{Polar---spectropolarimetry. \textsuperscript{4} Target} objects: \mbox{Deep---deep space} objects; NEO---near-Earth objects; LEO---low-Earth orbit. \textsuperscript{5} Year of commissioning of the instrument(s), where 202x denotes decision stage.
    }
\end{adjustwidth}
    \end{adjustwidth}
\end{table}

\subsection{NEO~Research}\label{sec:deepspace}
Recent advancements in FAI's optical monitoring facilities have solidified the institute's role as a key contributor to global planetary defence, marked by its membership in the International Asteroid Warning Network (IAWN). The~research into near-Earth objects (NEOs) employs a multi-faceted approach, combining direct observational campaigns with advanced characterisation techniques. These methods range from high-precision photometry of stellar occultations to determine asteroid physical properties to reflectance spectroscopy for investigating surface composition and taxonomy. The~significance of these capabilities is highlighted by contributions to major international efforts, such as the spectral analysis of the DART mission impact, and~is continually bolstered by the development of new, specialised~instrumentation.

\subsubsection{Asteroid Observation~Campaigns}\label{sec:asteroidoccultation}
In June 2023 FAI became a member of the International Asteroid Warning Network (IAWN).\endnote{International Asteroid Warning Network (IAWN): \url{https://iawn.net/index.shtml} (accessed on 15 July 2025).} This was made possible due to successful completion of modernisation works of optical monitoring facilities for  near-Earth space. In~the frame of this collaboration, FAI provided IAWN with the observation of the 99,942~Apophis asteroid, 450~m by 170~m in size, a~near-Earth and potentially hazardous object, obtained during the campaign in 2021, and~became a part of a global effort in establishing planetary defence~\cite{Reddy...2022PSJ.....3..123R}. Analysis of other asteroids and hazardous space objects observed by ATO instruments has also been reported in recent publications~\cite{Marshall...2022DPS....5451407M, Troianskyi...2024Icar..42016146T}.

Another valuable tool in asteroid research is asteroid occultation events. These are used for refining asteroid orbits and conducting detailed analyses of asteroid shape and size. It has also been shown that this technique can detect rings and satellites of asteroids. 
 The~efficiency of occultation event predictions and observations has increased significantly with the success of the GAIA mission~\cite{Tanga...2007AA...474.1015T}. Activity in this field of research is underway at the AstroHub at the ATO thanks to collaboration with the International Occultation Timing Association---East Asia (IOTA/EA).\endnote{International Occultation Timing Association---ast Asia (IOTA/EA): \url{https://www.perc.it-chiba.ac.jp/iota-ea/wp/about-iota-ea/} (accessed on 15 July 2025).} 

The first successful detection of an occultation event at the ATO was the occultation of star GAIA~DR3~903878551799227008 by asteroid (634) Hektor on 2023 December 20. The~event began at UTC~22$^h$44$^m$08$^s$.097 $\pm$ 0.45 s with a signal-to-noise ratio of S/N = 40, and~it finished at UTC~22$^h$44$^m$17$^s$.308 $\pm$ 0.45 s with S/N = 30. The~whole set of observations lasted from 22$^h$39$^m$55$^s$ to 22$^h$49$^m$04$^s$. The~star's brightness amplitude drop was 0$^m$.55. The~observation was carried out using the photometric channel of the AZT-20 in the first binning and an~exposure of 0.9 s, with~a CMOS~FLI~KL400 cooled down to $-$30 $^\circ$C. The~details can be found on the website.\endnote{Results of Asteroidal occultation. The~Summary of East Asia Results (Observation \#969. 624 Hektor): \url{http://hal-astro-lab.com/data/occult-e/occult-e.html} (accessed on 15 July 2025).}

We expect significant growth in asteroid observational activity with the ongoing installation of the new optical system, 70 cm Wide-Field-of-View Optical System (\mbox{WFOS-70}). 
To~develop the methodology for analysing the data expected from WFOS-70, a~series of observations of several selected star fields was conducted. For~each field, approximately $2000$~images with exposures ranging from $0.1$ to $0.2$~s were obtained. The~analysis and its results showed the fundamental possibility of detecting a signal at a level of $0.5^m$ or greater for objects brighter than $13^m$--$14^m$ due to the stellar magnitude drop in a moment of asteroid occultation. We expect that the WFOS-70 will be able to detect occultation events for even fainter~stars.

\subsubsection{On LEO Observation~Capabilities}\label{sec:indianlunartelescope}
The number of objects in Low-Earth Orbits (LEOs) has increased dramatically in the past decade~\cite{Bombardelli...2021JGCD...44..684B}. This is mainly due to the unfolding mega-constellation of satellites~\cite{Bonnal...2020AcAau.170..637B,Canoy...2024JSpRo..61..622C}. This number is expected to grow, as more players can launch such constellations soon, thanks to technological advancements and reduced launch prices. As~a result, there will be increased demands on various segments of Space Situational Awareness (SSA), including ground-based Space Surveillance and Tracking (SST), to~track LEO and provide updated information regarding LEO's whereabouts and direction of movement~\cite{Sorge...2025AcAau.229..600S}.
 There is also a growing need for re-entry prediction to ensure safety for ground structures, commercial airlines, and~other critical infrastructure~\cite{Lawrence...2022NatAs...6..428L}.

To enhance the capability of national SSA, particularly SST, which is also being developed at the ATO, new Wide-Field-of-View Optical Systems were developed and installed: WFOS-40, with~an aperture of 40 cm, and~WFOS-70, with~an aperture of 70 cm. These systems are equipped with direct-drive mounts that can slew at speeds of up to \mbox{50 degrees} per second, featuring controllers and software capable of conducting autonomous observations that cover the entire hemisphere in one night and track LEO with high precision. The~power of WFOS-40 has been demonstrated in several cases, including tracking of the CHANDRAYAAN-3 (ISRO) mission on 29 July 2023, between~its last Earth-bound manoeuvre on 25 July and its trans-lunar injection on 31~July.
 
\subsubsection{Spectral Observation of~Asteroids}\label{sec:asteroidspectra}
The spectral range of optical transparency of the Earth's atmosphere is between $350$ and $1000$~nm. In~this spectral range, asteroids only reflect solar radiation. Spectral and geochemical studies of meteorites, which are most likely fragments of asteroids, show that the features of the reflectance spectra in the range of $\sim$$200$--$2500$~nm characterise the chemistry and mineralogy of the~asteroids.

The methodology of spectral observations and analysis of asteroid spectra is somewhat different from the spectroscopy of other astronomical objects. The~differences in spectroscopy of stars and asteroids stem from the fundamental differences between these objects: stars are sources of radiation, while asteroids are reflective bodies \citep{2023SoSyR..57...61B}. Difficulties in spectroscopic observations of asteroids are due to their relative movement, which differs from the sidereal motion of stars, variation in the illumination conditions for asteroids, and~changes related to their~rotation. 

All this required the development of a new spectral instrument for the AZT-20 telescope, as~well as the appropriate methodology for conducting observations and analysing the obtained data. A~new long-slit spectrograph was designed and manufactured by FAI and ``Astrotech'' engineers. It is based on Volume-Phased Holographic Grating (VPHG) from Wasatch Photonics with $360$, $1800$, and~$2400$~lines per mm. For~asteroid spectroscopy, $360$~lines per mm mode of VPHG is used, which covers a wavelength range of $400$--$800$~nm with a dispersion of $4$ \AA~per pixel using a $9$~arcsec slit width. For~the wavelength calibration a He-Ar-Ne lamp is used, and~LED--tungsten sources are used for flat fielding. Another important feature of the new spectrograph comes with its detector. For~better performance in high-readout mode and faint sources, the~EMCCD iXon-888 is used with the ability to reduce the effective readout noise of the CCD chip using an electron multiplication register. Here we present some recent and outstanding results obtained with this new~instrument.

\subsubsection{Participation in DART~Mission}\label{sec:dartmission}
The Double Asteroid Redirection Test (DART) \citep{2018P&SS..157..104C} was a mission devoted to technology development for redirecting an asteroid that poses a threat to the Earth \citep{2016P&SS..121...27C, 2021PSJ.....2..173R}. NASA launched the mission on 21 November 2021. On~26 September 2022, at~23:14 UT, the~probe, weighing $550$~kg, collided with the surface of Dimorphos, a~satellite of Didymos (65,803), with~a relative speed of $6.6$~km/s. As~a result, a~large amount of material from Dimorphos was released, the~brightness of the asteroid increased by three magnitudes, and~the orbital period changed by almost 33 min \citep{2022MNRAS.tmp.3047T, 2022PSJ.....3..206F, 2022RNAAS...6..186Z, 2022MNRAS.515.2178M}.

FAI participated in this experiment by conducting spectral observation during the impact event using a spectrograph mounted in the prime focus of AZT-20. The~\mbox{results~\cite{Shestakova...2022RNAAS...6..223S,2023Icar..40115595S}} indicated that the rare event of alkali metal emission was detected during the impact. Using the flux ratios of these emissions, the~relative abundance of atoms was estimated, and~good agreement with the abundance of these elements in the Solar System \citep{2009ARA&A..47..481A, 2013ApJ...774L..32M} was found. The~phenomenon observed during the DART probe impact on Dimorphos is resonant fluorescence, that is, the~re-emission of solar photons at a wavelength corresponding to the transition from the ground level of the atom of the \mbox{corresponding element.}
   
\subsubsection{Asteroid~Taxonomy}\label{sec:dartmission}
Asteroids are considered possible sources of extraterrestrial natural resources (see, e.g.,~Chapter~3 in~\cite{NRC...2010dpeneoshms.book.N,Lewis...1996mtsurftacp.book.L,Hein...2020AcAau.168..104H}), and their classification is becoming increasingly popular with the development of appropriate technologies for their extraction. Their classification (taxonomy) is based on certain spectral features of various minerals. Asteroids can exhibit differences in reflectance across the visible and infrared spectral ranges, corresponding to the presence of various minerals and chemical compounds. For~example, the~presence of certain bands may indicate the presence of silicates or carbonaceous materials \citep{1974Sci...186..352M, 1996Icar..119..202H, 2021Icar..35414034E}.

Asteroid taxonomy research is conducted at FAI in close collaboration with the Institute of Astronomy of the Russian Academy of Sciences (INASAN, Russia). A~growing spectroscopic database of asteroids has been accumulated from ground-based observations carried out at the ATO during 2023--2024 in the spectral range $4000$--$7500$~\AA ~and by INASAN during 2013--2017 in the range $4000$--$9500$~\AA. In~total, $25$~asteroids were investigated, and~their taxonomy has been performed so far using the template method~\citep{2022INASR...7..143S}.
  
\subsubsection{Spectral Observation of~GEO}\label{sec:goespectra}
The spectrophotometric characteristics of Geostationary Equatorial Orbit (GEO), such as light curves, colour indices, and~reflective spectra, play a key role in developing identification methods, diagnosing satellite conditions, and~studying the influence of outer space on the properties of the satellite's materials \citep{2009ESASP.672E..41V, 2010amos.confE..47C, 2012amos.confE..14H, 2017AdSpR..59.2488V, 2025AdSpR..75.7365Z, 2021spde.confE.140Z}. For~this project, FAI utilises a spectrograph manufactured using volume-phase holographic gratings with $360$~lines \mbox{per mm} (R = 600) with a dispersion of $ 4.25$~\AA~ per pixel and installed in the prime focus of the AZT-20. This configuration enables the acquisition of GEO spectra with exposure times as short as $2$~s and a sufficient signal-to-noise ratio. In~total, reflective spectra for $48$~GEOs were obtained and analysed. This helped us to develop a preliminary methodology to distinguish the satellites by their bus, shape, and~age. 

\subsection{Deep Space~Research}\label{sec:deepspace}
Deep space research at the ATO is driven by international collaboration, tackling fundamental astrophysical phenomena from stellar evolution to the most energetic events in the universe. Significant partnerships focus on the rapid response capabilities of dedicated ATO telescopes for the follow-up observation of optical afterglows from Gamma-Ray Bursts (GRBs). This is complemented by developing novel methods to probe dust cloud morphology around White Dwarfs. Through these combined efforts, the~instruments at Assy-Turgen provide critical insights into the physics of cosmic explosions and \mbox{circumstellar environments.}

\subsubsection{Morphology of Dust Around White~Dwarfs}\label{sec:whitedwarfs}
As an illustration of international collaboration, the~participation of FAI in the Whole Earth Telescope campaign is noteworthy~\cite{vonHippel...2024ApJ...963..113V}.

This initiative focused on photometric observations and light curve analysis of the White Dwarf (WD) G29-38 within the $0.7$~to~$2.5~\upmu$m wavelength range. The~primary objective was to develop a methodology for determining the potential geometric characteristics of the dust cloud surrounding G29-38, utilising the eigenmodes of the WD's oscillations. These oscillations are observed in the visible spectrum and serve as a source of variable radiation flux for the dust, which primarily re-radiates in the infrared range. A~comprehensive study of the dust's morphology around WDs can illuminate its formation mechanisms, thereby providing additional insights into the progenitor bodies that contributed to its creation. Currently, there are no reliable empirical methods to determine the distribution and geometry of circumstellar matter precisely. G29-38 is an ideal testbed for these purposes, as~it not only exhibits signs of dust but also displays ZZ Ceti-type~oscillations.

The core concept of this study is to leverage the oscillations of G29-38, which manifest distinctly for a ground-based observer and the surrounding dust. This distinction can then be exploited to investigate the dust's distribution and morphology by observing oscillations across the visible and infrared spectral ranges. Such a study was facilitated by the extensive photometric data bank in the visible range, accumulated by ground-based observatories during international campaigns, and~significant advancements in infrared detector~technology.

Analysis of the oscillation mode characteristics and their behaviour across different wavelength ranges led to the following conclusion: the dominant response to the WD oscillations in near-infrared light is most likely attributable to the isotropic component of the combined oscillation modes. Unfortunately, the~isotropic components of pulsations are unsuitable for determining the geometric structure of the dust distribution. To~investigate dust morphology, it is imperative to measure its response for modes with $\ell \geq 1$, i.e.,~in the fundamental modes and potentially the non-isotropic components of the combined oscillation modes. This will necessitate meticulous future modelling. Furthermore, to~ascertain the proportion of infrared pulsations originating from the dust versus the stellar photosphere, detailed modelling of atmospheric pulsation modes is required, which depends on the efficiency of mode~identification.

An alternative approach would involve observations at wavelengths beyond $5~\upmu$m, where the spectrum of G29-38 is almost entirely dominated by emission from heated dust. Due to the variable nature of the WD pulsation modes, only low-amplitude pulsations, insufficient for our task, are observed at certain epochs. However, whenever G29-38 exhibits distinct pulsation frequencies, this could offer a novel diagnostic tool. Moreover, the~dust distribution itself may evolve. Changes in pulsation phase also warrant further investigation and may aid in exploring the geometry of the dust distribution and stellar surface oscillations. Although~not the primary objective of these studies, our analysis of the combined modes nevertheless indicates that all observed fundamental oscillation modes possess the same $\ell$ value. This finding may prove beneficial in future endeavours to identify the pulsation modes of G29-38. The~upcoming campaign targeting G29-38 is scheduled for the last quarter of~2025.

\subsubsection{Optical Afterglow of Gamma-Ray~Bursts}\label{sec:grbafterglow}
There are follow-up observations for the optical afterglow of Gamma-Ray Bursts (GRBs), which are the prompt gamma-ray releases caused by a compact binary merger or massive star gravitational collapse, outshining its hosting galaxy, and~observable later in the entire electromagnetic spectrum~\cite{Zhang...2018pgrb.book.....Z,Vigliano...2024Univ...10...57V}. Both newly installed and enhanced telescopes are made to enable conducting alert follow-up observations after a GRB~trigger. 

\textls[-23]{To date, the~most active ATO instruments in this field of research are \mbox{NUTTelA-TAO \citep{Maksut...2021RMxAC..53..169M, Grossan...2022SPIE12184E..8AG}}} and AZT-20 telescopes. NUTTelA-TAO is designed for the follow-up of the earliest possible optical GRB spectra to make a clear conclusion on the initial stages of the burst. More than $30$~follow-up observation telegrams have been published from this telescope. 
And~it is capable of catching the very early phase of fast-decaying optical afterglow, co-existing with the prompt $\gamma$-ray emission of the GRB~201015A \citep{Komesh...2023MNRAS.520.6104K}. 
Some greater statistics on photometric follow-up of GRB afterglow are shown by AZT-20. It dedicates part of the nighttime to work in ``alert'' mode, interrupting and rescheduling all lower-priority observations. Perspective usage and AZT-20's full potential will come with the methodology of spectral observations of the optical afterglow of GRBs associated with supernova~explosions.

\section{Future Prospects of the~AstroHub}\label{sec:futureprospects}
\subsection{Development of the Software Platform for~NEO}\label{sec:futuresoftware}
\subsubsection*{The Orbit Quality Score Method for the Ground-Based Observation~Strategy}\label{sec:futureorbitqualityscore}
Within the AstroHub framework, the~\emph{unweighted} Orbit Quality Score (OrQS) will be adopted as a site-level metric to assess how effectively an observatory maintains observational custody of objects across Low-Earth Orbit (LEO), Medium-Earth Orbit (MEO), Geostationary Orbit (GEO), and~cis-lunar/interplanetary regimes. OrQS is observer-agnostic; we formulate a novel method \citep{Akniyazov...2025IAC} for ground-based application to quantify the performance of a specific site and configuration in a way that is interoperable across clients and integration~levels.

\textls[-15]{\textit{What~OrQS~measures~and~why.} 
 Operationally, an~observatory is valuable when it \mbox{can (i) \emph{see}} an object often enough within a planning window to maintain custody and~(ii) balance \emph{first detections} with \emph{revisits} that stabilise orbit knowledge and tracking continuity~\citep{Schildknecht...1994GGAS...49.....S,Coder...2016AcAau.128..669C}. OrQS encodes these requirements with three transparent components that can be computed for each object without model-dependent weighting. The~score is interpretable, bounded, and~comparable between instruments and sites. This simplicity is deliberate: AstroHub needs a metric that is robust to heterogeneous catalogues, uneven visibility, and~mixed integration levels and~that supports rapid trades and schedule generation without heavy~tuning.}

\textit{Definitions.} For each object over a discretised observation window of $N$ time~bins, the following hold:
\begin{itemize}
    \item $T$ is the number of bins in which the object is actually observed (passes all site/instrument constraints);
    \item $C$ and $R$ are, respectively, the~counts of \emph{complementary} (first) detections and \emph{redundant} (resighting) detections within those $T$ bins; by construction $C+R=T$ when $T>0$;
    \item $P\in[0,1]$ is the time coverage fraction, $P=T/N$.
\end{itemize}

\textit{Per-object score.} The unweighted score is
\begin{equation}
    \label{eq:orqs_obj}
    \mathrm{OrQS}_{\mathrm{obj}} \;=\; \frac{P \;+\; \frac{C}{T} \;+\; \frac{R}{T}}{3},
    \qquad
\end{equation}
with ${\rm OrQS_{obj}}=0$ if $T=0$. Since $C+R=T$, for~$T>0$ this collapses to
\begin{equation}
    \label{eq:orqs_collapse}
    \mathrm{OrQS}_{\mathrm{obj}} \;=\; \frac{P+1}{3}.
\end{equation}

\textit{Logic of the formulation.} 
\begin{itemize}
    \item \emph{Coverage $P$} captures for how much of the window the site actually observes the object; weather, horizon masking, sky brightness, phase angle, and~tracking limits are all reflected through $T/N$.
    \item \emph{Normalising by $T$} for $C$ and $R$ asks the following: \emph{of the times the object could be seen, what fraction were first vs.\ repeat detections?} This treats short-visibility and long-visibility objects fairly; scarcity is already penalised via $P$.
    \item \emph{Collapse property.} Because $\tfrac{C}{T}+\tfrac{R}{T}=1$, the~unweighted score reduces to $(P+1)/3$. This yields a bounded, interpretable range: for $T>0$, $\mathrm{OrQS}_{\mathrm{obj}}\in\left[\tfrac{1}{3},\tfrac{2}{3}\right]$, reaching $\tfrac{2}{3}$ when $P=1$. The~constant ``$+1$'' term reflects the fact that, once observable, the~mix of first vs.\ repeat detections is fully accounted for within the observed time.
\end{itemize}

\textit{Ground-based interpretation.}
Although OrQS originated for space-based custody, the~formulation applies directly on the ground. The~coverage term $P=T/N$ naturally encodes local constraints (horizon, seeing, moonlight/sky brightness, weather downtime, trailing, and tracking/blur limits), so $\mathrm{OrQS}_{\mathrm{obj}}=(P+1)/3$ measures how effectively a site can maintain custody of each object within the window. Aggregating $\mathrm{OrQS}_{\mathrm{obj}}$ over representative catalogues provides a site-level figure of merit for comparing locations, planning upgrades, and~selecting~schedules.

\textit{Contribution to the AstroHub framework.} The OrQS method provides a compact, comparable metric~that achieves the following: 
\begin{itemize}
    \item Prioritises observation tasks by ranking objects and time bins via $\mathrm{OrQS}_{\mathrm{obj}}$;
    \item Compares instrument configurations (aperture, FOV, exposure, and tracking policy) using aggregated OrQS across target sets;
    \item Informs scheduling by highlighting bins with high expected coverage $P$ and by quantifying trade-offs between alert, follow-up, and~program time;
    \item Standardises reporting across clients at different integration levels with a single, observer-agnostic score.
\end{itemize}

\textit{Method outline.} In order to introduce the method, one should implement the \linebreak  following~algorithm:
\begin{enumerate}
    \item Discretise the simulation into $N$ time bins; apply line-of-sight, photometric, and~tracking constraints per bin.
    \item For each object, mark observable bins and set $T$.
    \item If $T=0$, set $\mathrm{OrQS}_{\mathrm{obj}}=0$ and proceed to the next object.
    \item Within the $T$ observable bins, label first detections as complementary ($C$) and revisits as redundant ($R$), noting $C+R=T$.
    \item Compute $P=T/N$ and evaluate \eqref{eq:orqs_obj} (equivalently \eqref{eq:orqs_collapse}) for the per-object score.
    \item Aggregate scores by orbit class and instrument configuration (e.g., distributions and means/medians) to support planning and design.
\end{enumerate}

\textls[-15]{\textit{Use case for Assy-Turgen Observatory AstroHub}. Specifically for the observatory needs, OrQS guides the following: (i) trade-offs between wide-FOV survey systems and narrow-FOV trackers for LEO, GEO, and~NEO samples via the distribution of $\mathrm{OrQS}_{\mathrm{obj}}$; (ii) schedule selection by favouring time blocks with larger $P$; and (iii) interoperable tasking and reporting with partially/fully integrated clients using a single, comparable site-level metric. In~practice, OrQS maps (sky/temporal heatmaps) derived from the same discretisation are used to steer pointing patterns and hand-offs between instruments while preserving~custody.}

As one of the key future developments for NEO research at the ATO, the~AstroHub will implement the OrQS method, a~standardised, site-level metric designed to transparently assess the effectiveness of maintaining observational custody of space objects across different orbital regimes. This robust, observer-agnostic score will be used to inform observation strategies, evaluate different instrument configurations, and~provide a common performance benchmark for all participating clients, regardless of their integration~level.

\subsection{Collaborations}\label{sec:futurecollaborations}
\subsubsection{Developing SSA~Capabilities}\label{sec:ariannegroup}
The number of artificial objects, including space resident objects (RSOs), is increasing steadily. According to Celestrack, as~of 15 July 2025 there are 30,425~objects in near-Earth orbits, and~at least $15,254$~of them are space debris.\endnote{Celestrack's satellites catalogues with orbital data: \url{https://celestrak.org/satcat/boxscore.php} (accessed on 15 July 2025).} The total number of objects has increased by $\sim$200 since April 2025. Refining orbital elements for such a large number of objects is a cumbersome task for any local-scale SSA system. Even such developed SSA systems as the United States Space Surveillance Network (US SSN) and the Russian Space Surveillance System (RSSS) experience challenges in keeping their catalogues \mbox{regularly updated.}

Several initiatives have been proposed to alleviate the problems related to safety and sustainable development in space, such as Kazakhstan's recent proposal to establish a Regional Center for Space Situational Awareness (RCSSA) based on Kazakhstan’s observatories under
the auspices of the United Nations at the Scientific and Technical Subcommittee of the COPUOS. To~move forward with this proposal, FAI updated the infrastructure of the ATO, started to build several roll-off domes to host small-aperture telescopes, and~signed an agreement with local business partners, the~``Astrotech'',~\endnote{AstroTech: \url{https://astrotech.kz/} (accessed on 15 July 2025).} to develop and install several ``shell'' domes for FAI's and partners' optical systems. Under~the aegis of RCSSA development and the~AstroHub initiative and~within the framework of the Telescopes Network project, FAI is currently working on agreements with partners from India, China, France, and~Poland to install their SSA-specific instruments at the ATO, see Table~\ref{tab:assyhubclientsplans}. It is expected that the number of optical and electronic systems at the ATO for SSA needs will increase up to dozens in a few~years.

\subsubsection{Pulkovo~Observatory}\label{sec:pulkovo}
In the summer of 2022, a~40 cm Meade LX-200 telescope was installed at the ATO and received its first light as part of a collaboration between FAI and the Main (Pulkovo) Astronomical Observatory of the Russian Academy of Sciences (Saint Petersburg, Russia). Previously, at~another observatory, it has been used for photometric monitoring of young stars, blazars, and~active galactic nuclei (AGN). These objects will remain the core of the observational program of the LX-200 telescope after its movement to the ATO. Additional on-site tests have shown that it is advisable to add astrometric observations of planetary satellites and asteroids to the photometric part of the program~\cite{Bikulova...2021P&SS..20405245B}. Transit timing variation (TTV) observations of exoplanetary transits should become a separate direction in the observational~program. 

Despite the relatively small size of the LX-200 telescope, good astroclimatic conditions allowed single shots with a 30~s exposure to accumulate sufficient signal for stars up to $18^m$, and~when combining a series of shots, stars up to $20^m$ were detected with~a measurement accuracy of $0.02^m$. 
Astrometric measurements of asteroid coordinates in images taken with this instrument demonstrated an accuracy of $0.04''$--$0.1''$.

The telescope is participating in the asteroid and comet hazard warning program; the observation results are sent to the Minor Planet Center (MPC). To~date, regular remote sensing observations have been conducted on LX-200 \citep{Khovritchev...2023PPulO.231....9K}, particularly astrometric observations of the satellites of Jupiter and Saturn, as~well as several near-Earth asteroids \mbox{(1998 HH49,} 1998 KV2, 1998 QK28, 2004 OT11, 2005 XY7, 2007 PE8, and 2007 UM12). 

Joining the AstroHub became beneficial for Pulkovo scientists since there was the possibility to transfer part of their observational tasks to the AZT-20 telescope and other FAI instruments. With~astrometric accuracy better than 20 mas on AZT-20, it is possible to measure the relative radial velocities of the components in visual binary stars, objects from the Pulkovo science program. It has been shown that an accuracy of better than 300 m per second is achievable~\cite{Khovritchev...2022PPulO.227...58K}.

\section{Conclusive~Remarks}\label{sec:astrohubconclusion}
The AstroHub concept is likely the right case when \textit{``the whole is different than the sum of its parts''}.\endnote{The philosophical roots of the idea that ``the whole is something besides (or more than) the sum of its parts'' trace back to Aristotle, the~ancient Greek philosopher. He emphasised that the arrangement, form, and~essence of a whole give it properties not found in its constituent parts alone. This concept is fundamental to the idea of synergy, where combined action produces a total effect that is greater than (or different from) the sum of its individual effects. Similar conclusions, though~in a slightly different context, were also made in Holism and Gestalt psychology.} With that, the~AstroHub is about integration and network effects. It is a platform model from the beginning. It recognises that the value is not just in the physical location but~in the connections between the~participants.

In this mini-review we were also challenged with summarising the key achievements in developing the Assy-Turgen Astronomical Hub. We attempted to show its significance as a growing centre for astronomical research and instrumentation in Kazakhstan and Central Asia. We have provided a general conceptual view on the future of the observatory and illustrated its potential impact on the highlighted research~fields. 

When we propose joining the AstroHub, we are offering entry into a dynamic, flexible, and~powerful scientific ecosystem, with~a clear and transparent~framework.

\vspace{6pt}
\authorcontributions{Conceptualisation, C.O. and A.K.; methodology, C.O., A.S., D.Y., M.K., M.M., S.S., C.A., V.K. and Y.A.; software, A.S., D.Y., M.M. and V.K.; validation, S.S., M.M., V.K. and Y.A.; formal analysis, G.A., R.V. and R.K.; investigation, A.S., G.A., M.K., S.S. and V.K.; resources and supervision, C.O. and M.K.; data curation, S.S., M.M., C.A. and V.K.; writing---original draft preparation, and writing---review and editing, C.O., A.S., D.Y., M.M., S.S., A.K., C.A., V.K. and Y.A.; visualisation, A.S., D.Y., C.A., V.K. and Y.A.; project administration, C.O., R.V. and R.K.; funding acquisition, C.O. All authors have read and agreed to the published version of the manuscript.}

\funding{This research is funded by the Science Committee of the Ministry of Science and Higher Education of the Republic of Kazakhstan (Grant~No.~BR21881880, Grant~No.~BR24992807, and Grant~No.~BR24992759).}

\institutionalreview{Not applicable.}

\informedconsent{Not applicable.}

\dataavailability{No new data were produced within this research.}

\acknowledgments{We would like to thank all the organisers of the ``Hot Stars 2024'' conference. The authors express their gratitude to the editors and anonymous referees for their valuable comments that improved the presentation of the~results.}

\conflictsofinterest{The authors declare no conflicts of interest. The~funders had no role in the design of the study; in the collection, analyses, or~interpretation of data; in the writing of the manuscript; or in the decision to publish the~results.}

\abbreviations{Abbreviations}{
The following abbreviations are used in the manuscript:

\begin{longtable}[l]{@{}ll}
    ATO & Academician Omarov Assy-Turgen~Observatory\\
    AZT-20 & Astronomical Reflecting Telescope with 1.5 m~aperture\\
    \textls[-25]{CDK-700 (NUTTelA-TAO)} & Nazarbayev University Transient Telescope~with  70 cm~aperture\\
    DART & Double Asteroid Redirection~Test\\
    DIMM & Differential Image Motion~Monitor\\
    FAI & Fesenkov Astrophysical~Institute\\
    FAIR & \textls[-20]{Findable, Accessible, Interoperable, Reusable (principles)}\\
    GEO & Geostationary Earth~Orbit\\
    GRB & Gamma-Ray~Burst\\
    IOTA/EA & \textls[-20]{International Occultation Timing Association---East Asia}\\ 
    INASAN & Institute of Astronomy of the Russian~Academy of~Sciences\\
    ISRO & Indian Space Research~Organisation\\
    IVOA & International Virtual Observatory~Alliance\\
    KPO & Kamenskoye Plateau~Observatory\\
    LEO & Low-Earth~Orbit\\
    MEO & Medium-Earth~Orbit\\
    MCDA & Multi-Criteria Decision~Analysis\\
    NEO & Near-Earth~Object\\
    OrQS & Orbit Quality Score (method)\\
    RC-500 & Ritchey--Chretien optical telescope with 50~cm aperture\\
    RSSA & Regional Space Sustainability~Agreement\\
    SSA & Space Situational~Awareness\\
    SST & Space Surveillance and~Tracking\\
    SROs & Space Resident~Objects\\
    TSHAO & Tien-Shan Astronomical~Observatory\\
    VPHG & Volume-Phased Holographic~Grating\\
    WFOS-40 & Wide-Field-of-View Optical System with 40~cm aperture\\
    WFOS-70 & Wide-Field-of-View Optical System with 70~cm aperture\\
    Zeiss-1000 & Carl Zeiss Optical Telescope with 1~m aperture\\
    Zeiss-800M & Modernised Carl Zeiss Optical Telescope~with  80~cm aperture\\
\end{longtable}}

\appendixtitles{yes} 
\appendixstart
\appendix
\section{Telescope Hosting~Scheme}\label{sec:telescopehostingscheme}
Based on common practices at observatories worldwide, there is a breakdown of how telescopes hosting observatories is usually managed. These stages include the following:
\begin{enumerate}[label=,leftmargin=*,labelsep=0mm]
    \item[]\textbf{The Initial Proposal and Scientific Case.} It almost always begins with a formal proposal from the prospective client (the telescope owner) to the host observatory's management. This proposal typically needs to address the following: 1. Scientific Justification: What is the scientific purpose of the telescope? What unique research will it conduct? 2. Technical Specifications: A detailed description of the telescope, mount, camera, and~all auxiliary equipment. This includes its physical size, weight, power requirements, and~operational software. 3. Benefit to the Host (if any): For professional collaborations, the~proposal might outline benefits to the host observatory, such as offering a percentage of observing time to the host's community, sharing data, or~collaborating on research.
    \item[]\textbf{The Hosting Agreement (The Contract).} Once a proposal is accepted in principle, a~formal hosting agreement is negotiated and signed. This is the central legal document that governs the entire relationship. Key clauses in a typical agreement include the following.
    
    A. Definition of Services and Infrastructure Provided by the Host, namely the following: 
    \begin{enumerate}
        \item Physical Space: A specific pier or pad within an existing dome or a plot of land for constructing a new pavilion.
        \item \textls[-15]{Infrastructure: Guaranteed access to reliable electricity (often with UPS backup), high-speed Internet (specifying bandwidth), and~physical security \mbox{(fencing, surveillance).}}
        \item On-Site Support: This is a critical point of the contract. It can range from basic ``emergency hands'' support (an operator physically checking on the equipment if it fails) to more comprehensive technical support for installation and maintenance.
        \item Environmental Monitoring: Access to the observatory's real-time weather data (seeing, wind speed, and humidity) is standard, which is crucial for \mbox{robotic operation.}
    \end{enumerate}
    B. Responsibilities of the Client (Telescope Owner): This outlines the client's obligations, namely the following:
    \begin{enumerate}
        \item Equipment: The client is responsible for providing, shipping, and~insuring their own fully functional and tested telescope system. Many hosting sites emphasise that equipment should be thoroughly tested for remote operation before it arrives.
        \item Installation and Decommissioning: The client is responsible for the costs and logistics of installing the equipment. Crucially, the~agreement will state that the client is also responsible for removing all equipment and clearing the site at the end of the contract term.
        \item Compliance: The client and their personnel must adhere to all of the host observatory's site safety regulations, operational procedures, and~policies.
    \end{enumerate}
    C. Financial Terms~include the following:
    \begin{enumerate}
        \item Hosting Fee: This is the core cost. It can be structured as a recurring monthly or annual fee. The~price often depends on the size of the telescope and the level of support required. For~purely commercial hosting, this is the primary \mbox{financial transaction.}
        \item Additional Costs: The agreement will specify costs for any services beyond the basic package, such as dedicated technical support (billed at an hourly rate), use of machine shops, or~specialised shipping and handling.
    \end{enumerate}
    D. Data Rights and Ownership. This is a fundamental clause, especially for professional instruments. It~regulates the following:
    \begin{enumerate}
        \item Default Position: Typically, the~data collected by a client's telescope belongs exclusively to the client. The~host observatory has no inherent rights to it.
        \item Negotiated Sharing: In collaborative or scientific hosting agreements, this clause would be heavily modified. It would detail the specifics of data sharing, proprietary periods on how long the client has exclusive access before data becomes public or shared, and~publication policies.
    \end{enumerate}
    E. Term and Termination. The~agreement specifies the duration of the hosting period. It outlines the conditions under which either party can terminate the agreement, including notice periods and responsibilities for~decommissioning.

    \item[]\textbf{Installation and Commissioning.} This is the logistical phase where the client's plans are physically realised. There are few stages to maintain:

    \begin{enumerate}
        \item Pre-Installation Checks: The host observatory staff will review the client's final installation plan to ensure it is safe and compatible with the site infrastructure.
        \item On-Site Work: The client's team travels to the observatory to install the telescope. The~host's role here is defined by the agreement---it could be as simple as providing access and power, or~it could involve active assistance from staff astronomers \mbox{and technicians.}
        \item Commissioning and Checkout: Once installed, the~telescope undergoes a commissioning phase to ensure it is operating correctly in its new environment. This includes pointing tests, focus calibration, and~remote connectivity checks and~sometimes can include very formalised procedures.
    \end{enumerate}

    \item[]\textbf{Routine Operations.} Once operational, the~management of the telescope depends on the agreement model and can consist of the~following:
    \begin{enumerate}
        \item Remote Operation. The~vast majority of hosted telescopes are operated remotely by the client via the Internet. The~client is responsible for scheduling observations, monitoring the system, and~retrieving their data.
        \item Fault Resolution. When something goes wrong, the~process is dictated by the agreed-upon support level. The~client first attempts to diagnose and fix the issue remotely. If~physical intervention is needed, the~client might use on-site cameras to inspect the equipment. If~hands-on work is required, they will contact the host observatory's on-site support team, which would then be governed by the terms of the \mbox{support contract.}
    \end{enumerate}
\end{enumerate}

\section{Patents}\label{sec:patents}
There are patents resulting from the works reported in this manuscript.

\subsection{Star Finder. Astroclimate~Monitoring}
``Star Finder. Astroclimate monitoring''. Certificate of Copyright registered by the National Institute of Intellectual Property of the Republic of Kazakhstan, No.~54042, dated 30 January~2025.

The program is designed to control the process of shooting sources on a DIMM device, as~well as to calculate astronomical seeing parameters based on statistical analysis of image fluctuations caused by atmospheric~turbulence.

\subsection{Planner for Astronomical~Observations}
``Planner for astronomical observations''. Certificate of Copyright registered by the National Institute of Intellectual Property of the Republic of Kazakhstan, No.~43289, dated 28 February~2024.

The program is designed to optimise the distribution of time for conducting astronomical observations on telescopes, taking into account the observatory's location and the date of observations. The~program takes into account twilight, the~moments of rise and set of objects, the~time of their visibility above the horizon, the~influence of lunar illumination, and the maximum height of objects at the moment of~culmination.

\subsection{Astronomical~Calendar}
``Astronomical calendar''. Certificate of Copyright registered by the National Institute of Intellectual Property of the Republic of Kazakhstan, No.~40376, dated 10 November~2023.

The program is designed to perform astronomical calculations to determine the time of sunrise and sunset, the~length of daylight hours, the~start and end time of civil, navigational, and astronomical twilight, the~rise and set of the Moon and its phases, stellar time, and the~time of the autumnal and vernal equinoxes and the~summer and winter~solstices.

\begin{adjustwidth}{-\extralength}{0cm}
\printendnotes[custom] 

\reftitle{References}

\PublishersNote{}

\end{adjustwidth}

\end{document}